\documentclass[fleqn,usenatbib]{mnras}

\usepackage{newtxtext,newtxmath}
\usepackage[T1]{fontenc}

\DeclareRobustCommand{\VAN}[3]{#2}
\let\VANthebibliography\thebibliography
\def\thebibliography{\DeclareRobustCommand{\VAN}[3]{##3}\VANthebibliography}

\usepackage{graphicx}

\usepackage{amsmath}
\usepackage{amssymb}

\newcommand{\vpeak}{V_{\rm peak}}
\newcommand{\vmax}{V_{\rm max}}
\newcommand{\mpeak}{M_{\rm peak}}
\newcommand{\minfall}{M_{\rm infall}}

\newcommand{\sigmaLogM}{\sigma_{\rm logM}} 
\newcommand{\Mone}{M_{\rm 1}}
\newcommand{\Mmin}{M_{\rm min}} 
\newcommand{\Mcut}{M_{\rm cut}} 

\newcommand{\hMsun}{ h^{-1}{\rm M_{ \odot}}}
\newcommand{\hMpc}{ h^{-1}{\rm Mpc}}

\newcommand{\ihMpcC}{ h^{3}{\rm Mpc}^{-3}}

\newcommand{\ihMpc}{ h\,{\rm Mpc}^{-1}}

\newcommand{\sig}{\sigma_{8}}
\newcommand{\OmM}{\Omega_\mathrm{M}}
\newcommand{\Omb}{\Omega_{\rm b}}
\newcommand{\h}{h}
\newcommand{\ns}{{n_{\rm s}}}
\newcommand{\Mnu}{M_{\rm \nu}}
\newcommand{\wa}{w_{\rm a}}
\newcommand{\wz}{w_{0}}

\defcitealias{C17}{C17}

\title[On the origin of the evolution of the HOD]{On the origin of the evolution of the halo occupation distribution}

\author[S. Contreras et al.]{
Sergio Contreras,$^{1}$\thanks{E-mail: sergio.contreras@dipc.org }
Idit Zehavi$^{2}$\thanks{E-mail: idit.zehavi@case.edu}
\\
$^{1}$ Donostia International Physics Center, Manuel Lardizabal Ibilbidea, 4, 20018 Donostia, Gipuzkoa, Spain.\\
$^{2}$ Department of Physics, Case Western Reserve University, Cleveland, OH 44106, USA.\\
}

\date{Accepted XXX. Received YYY; in original form ZZZ}

\pubyear{2023}

\begin{document}
\label{firstpage}
\pagerange{\pageref{firstpage}--\pageref{lastpage}}
\maketitle

\begin{abstract}
We use the TNG300 magneto-hydrodynamic simulation and mock catalogues built using subhalo abundance matching (SHAM) to study the origin of the redshift evolution of the halo occupation distribution (HOD). We analyse stellar-mass selected galaxy samples with fixed number densities, spanning the redshift range $0 \le z \le 3$. We measure their halo occupation functions and fit the HOD parameters to study their evolution over cosmic time. The TNG300 galaxy population strongly depends on the baryonic physics implemented in the simulation. In contrast, the galaxy population predicted by a basic SHAM model without scatter is a direct result of the cosmology of the dark matter simulation. We find that the HOD evolution is similar for both models and is consistent with a previous study of the HOD evolution in semi-analytical models. Specifically, this is the case for the ratio between the characteristic halo masses for hosting central and satellite galaxies. The only HOD parameter whose evolution varies across models is $\sigmaLogM$, which contains information about the stellar mass-halo mass relation of the galaxies but does not strongly impact galaxy clustering. We also demonstrate that the dependence on the specific values of the cosmological parameters is small. We conclude that the cosmology of the galaxy sample, i.e. the cosmological hierarchical growth of structure, and not the baryonic physics prescriptions, governs the evolution of the HOD for stellar mass-selected samples. These results have important implications for populating simulated lightcones with galaxies and can facilitate the interpretation of clustering data at different redshifts.    
\end{abstract}


\section{Introduction}
\label{sec:intro}
In the standard picture of hierarchical structure formation, galaxies reside in dark matter haloes.
The formation and evolution of the haloes is dominated by gravity, with the haloes growing by accretion and mergers.
The formation of the galaxies and their relation to the dark matter haloes is more complex and depends on the detailed physical processes, leading to the varied observed galaxy properties. 
As the haloes merge and evolve, the haloes will often host more than one galaxy (since galaxy merger is a slower process). The evolution of the galaxies may thus be impacted by both the baryonic physics and by the merger history of their host haloes.

One of the most useful and popular ways to characterise the dark matter halo-galaxy connection is by measuring the average number of galaxies that populate haloes as a function of halo mass, which provides the basis for the halo occupation distribution (HOD) framework \citep{Jing:1998a, Benson:2000, Peacock:2000, Berlind:2003, Zheng:2005, Zheng:2007, Guo:2015a}.
The HOD formalism has been widely used to interpret clustering data (e.g., \citealt{Zehavi:2011, Guo:2015}), to characterise different galaxy populations (\citealt{C13, Yuan:2022a}), to create mock galaxy catalogues
(e.g., \citealt{Grieb:2016}), to examine galaxy assembly bias effects (e.g., \citealt{Zehavi:2018, Salcedo:2022}) or even constrain cosmological parameters (e.g., \citealt{Cacciato:2013, More:2015, AEMULUS3, Miyatake:2022, Yuan:2022b, AEMULUS5}).

The HOD model's strengths as a technique for creating mock galaxy catalogues include its ability to reproduce realistic galaxy clustering, its flexibility, and computational efficiency. Populating a dark matter simulation with galaxies, over large cosmological volumes,  takes mere seconds and requires only the position and mass of dark matter haloes. Some HOD improvements, such as velocity bias \citep{Guo:2015a} or assembly bias \citep{Hearin:2016, Xu:2020}, may also necessitate the simulation's dark matter particles or additional halo properties (see \citealt{Yuan:2021} for the latest developments on HOD modelling). These requirements are significantly smaller than those of other techniques, such as subhalo abundance matching (SHAM, \citealt{Vale:2006, Conroy:2006, Guo:2014,C15, Kulier:2015, ChavesMontero:2016, Lehmann:2017, C21a, C21c, Favole:2022, C23c, C23a}) or semi-analytical models of galaxy formation (SAMs, e.g., \citealt{Kauffmann:1993, Cole:1994, Bower:2006, Lagos:2008, Benson:2010, Benson:2012, Jiang:2014, Croton:2016, Lagos:2018, Stevens:2018, Henriques:2020}), which require higher resolution simulations, subhalo property computation, and, in most cases, halo merger trees. In turn, these requirements are smaller than those of hydrodynamic simulations, which are arguably the most advanced way we have today to model galaxies on cosmological volumes.

In hydrodynamic simulations (such as EAGLE, \citealt{Schaye:2015}; Illustris,\citealt{Illustrisa}; Magneticum, \citealt{Magneticum}; HorizonAGN, \citealt{HorizonAGN}; Simba, \citealt{SIMBA}; IllustrisTNG, \citealt{TNGa, TNG50a} and MillenniumTNG, \citealt{MTNGa}), dark matter particles are modelled alongside baryonic particles/cells. These simulations are then able to directly reproduce the interaction between dark matter and baryons and provide unique opportunities to study, in detail, the evolution of galaxies in a cosmological context. The downside of these simulations is their high computational cost, which can be an order of magnitude larger than dark matter-only simulations. Hence, hydrodynamic simulations and SAMs are often used to enhance the accuracy of other, more practical, approaches for modelling galaxies, such as HODs and SHAMs.

Our work follows that of \citet[][\citetalias{C17} hereafter]{C17}, where we studied the evolution of the HODs of stellar mass-selected samples from two different semi-analytic models generated from the same dark matter simulation. In the SAMs, the haloes from the N-body simulations are populated with galaxies using analytical prescriptions for the baryonic processes. Following the dark matter merger trees, galaxies merge and evolve as new stars form and previous generations of stars change. The evolution of the HOD is characterised by fitting a parametric form to the HODs at different redshifts and studying the evolution of the fitting parameters. \citetalias{C17} present a simple evolution model for each of the HOD parameters. This evolution can be used to populate lightcones constructed from N-Body simulations (e.g., \citealt{Smith:2017, Smith:2022}) or for modelling clustering data at different redshifts. Although the HODs describing the two SAMs exhibit some differences, the evolution of HOD parameters followed a similar pattern. These findings may suggest that the evolution of the HOD is governed by cosmology and not galaxy formation physics.

In this paper, we explore the evolution of the HOD of stellar mass-selected samples for two distinct galaxy population models: a state-of-the-art hydrodynamical simulation, the TNG300, whose galaxy population strongly depends on the baryonic processes of the model, and a basic SHAM model without scatter. In the SHAM model, each subhalo in the simulation is assigned a stellar mass by assuming a monotonic relation to
a subhalo property ($\vpeak$, the peak value of the maximum circular velocity over the assembly history, in our case), such that the subhalo with the highest value of $\vpeak$ has the largest stellar mass and so on
(see \S~\ref{subsec:SHAM} for more details). Since we construct our galaxy samples based on a fixed number density, the galaxy samples produced by the SHAM model do not depend on any galaxy formation physics, but rather on the simulation's cosmology. We find that the HODs evolve nearly identically in both models, indicating that the evolution is determined by the cosmological hierarchical accretion picture and not by the galaxy formation physics. Having a universal way in which the HOD parameters evolve, independent of the galaxy formation model assumed, justifies some of the ansatzes assumed today when constructing simulated lightcone galaxy catalogues.

This paper is organised as follows. The simulations and galaxy population models used in this study are described in \S~\ref{sec:models}. The evolution of HOD in each of these models is depicted in \S~\ref{sec:HODev} and subsequently analysed \S~\ref{sec:discussion}. We conclude in \S~\ref{sec:Summary}.
Appendix~\ref{sec:EAGLE} presents results for the evolution of the HOD in the EAGLE hydrodynamical simulation.  Appendix~\ref{sec:cosmo_HODparam} examines the dependence on the values of the cosmological parameters. 
Unless otherwise stated, the standard units in this paper are $\hMsun$ for masses, $\hMpc$ for distances, $\rm km\ s^{-1}$ for the velocities, and all logarithm values are in base 10.

\section{Models of Galaxy Clustering} 
\label{sec:models}
In this section, we describe the galaxy population models employed in the construction and characterization of our galaxy samples. In \S~\ref{subsec:TNG300}, we introduce the TNG300 cosmological magneto-hydrodynamic simulation, as well as its dark matter-only counterpart, TNG300-Dark. In \S~\ref{subsec:SHAM}, we present the SHAM method employed to populate the TNG300-Dark. In \S~\ref{subsec:HOD}, we describe briefly the halo occupation distribution framework, used to characterise the different galaxy samples. Finally, in \S~\ref{subsec:nden}, we specify how we select and compare the galaxies from TNG300 and the SHAM mock.

\subsection{The TNG300}
\label{subsec:TNG300}
In this work, we use galaxy samples from the Illustris-TNG300 simulation (hereafter TNG300). This simulation is part of ``The Next Generation'' Illustris Simulation suite of magneto-hydrodynamic cosmological simulations (IllustrisTNG; \citealt{TNGa, TNGb, TNGc, TNGd, TNGe}), the successor of the original Illustris simulation \citep{Illustrisa,Illustrisb,Illustrisc,Illustrisd}. TNG300 is one of the largest publicly available high-resolution hydrodynamic simulations\footnote{\url{https://www.tng-project.org/}}. The simulated volume is a periodic box of 205 $\hMpc$ ($\sim300$ Mpc) aside and has 100 outputs between z = 127 and z = 0. The number of gas cells and dark matter particles is $2500^3$ each, implying a baryonic mass resolution of $7.44\times10^6\,\hMsun$ and a dark matter particle mass of $3.98\times 10^7\,\hMsun$. The cosmological parameters assumed in the simulations, $\OmM$ = 0.3089, $\Omb$ = 0.0486, $\sig$ = 0.8159, $\ns$ = 0.9667 and $h$ = 0.6774, are consistent with recent Planck values \citep{Planck2015}. 

TNG300 was run using the \texttt{AREPO} code \citep{AREPO} and features a number of enhancements over its predecessor, the Illustris simulation, including: an updated kinetic AGN feedback model for low accretion rates \citep{Weinberger:2017}; an improved parameterization of galactic winds \citep{Pillepich:2018}; and inclusion of magnetic fields based on ideal magneto-hydrodynamics \citep{Pakmor:2011,Pakmor:2013,Pakmor:2014}. 
The free parameters of the model were calibrated to ensure that the simulation agrees with a number of observations: (i) the stellar mass function, (ii) the stellar-to-halo mass relation, (iii) the total gas mass contained within the virial radius ($r_{500}$) of massive groups, (iv) the stellar mass – stellar size relation and the black hole-galaxy mass relation at $z = 0$, and (v) the overall shape of the cosmic star formation rate density up to $z \sim 10$. The TNG simulations also successfully reproduce many other observables not directly employed in the calibration process (e.g., \citealt{TNGb, TNGd, Springel:2018, Vogelsberger:2020}). 

To identify (sub)haloes/galaxies, a friends-of-friends group finder \citep[{\tt FOF};][]{Davis:1985} is first used to identify the haloes (with linking length 0.2), within which gravitationally bound substructures are then located and hierarchically characterised using the {\tt SUBFIND} algorithm \citep{Springel:2001}. The {\tt SUBFIND} catalogue contains both central and satellite subhaloes, with the position of the centrals coinciding with the {\tt FOF} centres (defined as the minimum of the gravitational potential).

We use as well the dark matter-only version of TNG300, which we refer to as TNG300-Dark. This simulation has the same initial conditions, volume, cosmology, number of outputs and number of dark matter particles as its hydrodynamic counterpart, but with a mass particle of  $4.27\times 10^7\,\hMsun$. We also utilize the merger trees of the simulation, run with the \texttt{SUBLINK} algorithm \citep{RodriguezGomez:2015}, to compute $\vpeak$ for the subhaloes, needed for constructing the SHAM catalogue.  

\subsection{The subhalo abundance matching}
\label{subsec:SHAM}

Subhalo abundance matching (e.g., \citealt{Vale:2006, Conroy:2006}) is an empirical method for populating the subhaloes of a $N$-body simulation with galaxies. In its most fundamental form, SHAM assumes a monotonic mapping between the (sub)halo mass of the central galaxies and their stellar mass or luminosity. Recent implementations of SHAM incorporate scatter and satellite galaxies by utilizing subhalo properties at infall or the maximum values throughout their assembly history. These modifications are necessary in order to obtain realistic clustering predictions.

One of the main advantages of SHAM approach is the computational efficiency and simplicity. In contrast to HOD models (described below in \S~\ref{subsec:HOD}), which have at minimum five free parameters, standard SHAM models use a single free parameter, the scatter between the subhalo property used and the stellar mass, in the majority of their implementations. Additionally, SHAM predicts galaxy clustering in rough accordance with hydrodynamical simulations and reproduces some, but not all, of its galaxy assembly bias signal (\citealt{ChavesMontero:2016}; see also \citealt{C21c, C23b}). At the same time, due to the necessity of identifying the subhaloes, the resolution of the N-body simulation required to run a SHAM is greater than for other galaxy clustering models. Furthermore, SHAM models typically require some subhalo properties that are not always provided by the N-body codes and need to be computed from the subhalo merger trees, such as the peak value of the subhalo over the assembly history ($\mpeak$), the peak value of the maximum circular velocity ($\vpeak$), or the subhalo mass at infall ($\minfall$).

In this paper, we create SHAM mocks with the subhalo property $\vpeak$ using the TNG300-Dark simulation. $\vpeak$ is defined as the peak value of $\vmax$ over the suhalo's evolution, where $\vmax$ is the maximum circular velocity ($\vmax \equiv {\rm max}(\sqrt{GM(<r)/r})$). It has been widely used as a SHAM primary property and has been shown to provide a tighter relation to the stellar mass of galaxies than other properties (see also the discussion in \citealt{Campbell:2018}). We use the stellar mass function of the TNG300 to assign stellar masses to the subhaloes. As we select galaxies based on number density, and use a SHAM without scatter, the choice of the stellar mass function has no impact on the results. 

We chose a SHAM without scatter created from the dark matter-only simulation since such a model is not influenced by galaxy formation physics and results purely from the input cosmology of the N-body simulation. This is in direct contrast to the case of a hydrodynamic simulation, where baryons are carefully modelled to create realistic galaxy population samples. For completeness, we also tested a SHAM model with scatter, which is in the middle of these two extremes, where the scatter is added to better mimic the properties of TNG300. However, as the results from this model were almost identical to the other two models, we chose not to include them here for the sake of clarity of presentation.

\subsection{The halo occupation distribution}
\label{subsec:HOD}

\subsubsection{HOD modelling}
\label{subsubsec:HODmodel}
The HOD formalism describes the relationship between galaxies and haloes in terms of the probability distribution that a halo of virial mass $M_{\rm h}$ contains $N$ galaxies of a particular type, as well as the spatial and velocity distributions of galaxies within haloes. The fundamental component is the halo occupation function, $\langle N(M_{\rm h}) \rangle$, which represents the mean number of galaxies as a function of halo mass. This approach has the advantage of not requiring assumptions about the physical processes that drive galaxy formation and can be empirically derived from observations. Additionally, by utilizing only the positions and masses of the dark matter haloes (rather than subhaloes) to populate with galaxies, the required resolution of the N-body simulation is significantly lower than for a SHAM.

Standard applications typically assume a cosmology and a parameterized form for the halo occupation functions, which are motivated by the predictions of SAMs and hydrodynamics simulations (e.g., \citealt{Zheng:2005}). The HOD parameters are then constrained using measurements of galaxy clustering from large surveys.
This method essentially converts galaxy clustering measurements into a physical relation between the galaxies and dark matter haloes, paving the way for comprehensive tests of galaxy formation models. 

An important application of this approach is the generation of simulated galaxy catalogues by populating dark matter haloes in N-body simulations with galaxies that reproduce the desired clustering. This method has gained popularity due to its low computational cost and high performance (e.g., \citealt{Manera:2015, Zheng:2016, Yuan:2021}). The halo occupation function is typically provided at a specific redshift or over a narrow redshift interval. To generate a mock galaxy catalogue over a wide range of redshifts (e.g., lightcone), an HOD model with a dependence on redshift may be needed. In \citetalias{C17}, we presented a novel approach to describe the HOD as a function of redshift. There, we studied the HOD evolution for stellar-mass selected galaxies since $z=3$, for two different SAMs. Even though the HODs of those two models were different, the evolution of their HODs was similar. A simplified version of our model was later used by \citet{Smith:2017, Smith:2022} to populate simulated lightcones built from N-body simulations to create more realistic galaxy catalogues.

\begin{figure}
\includegraphics[width=0.45\textwidth]{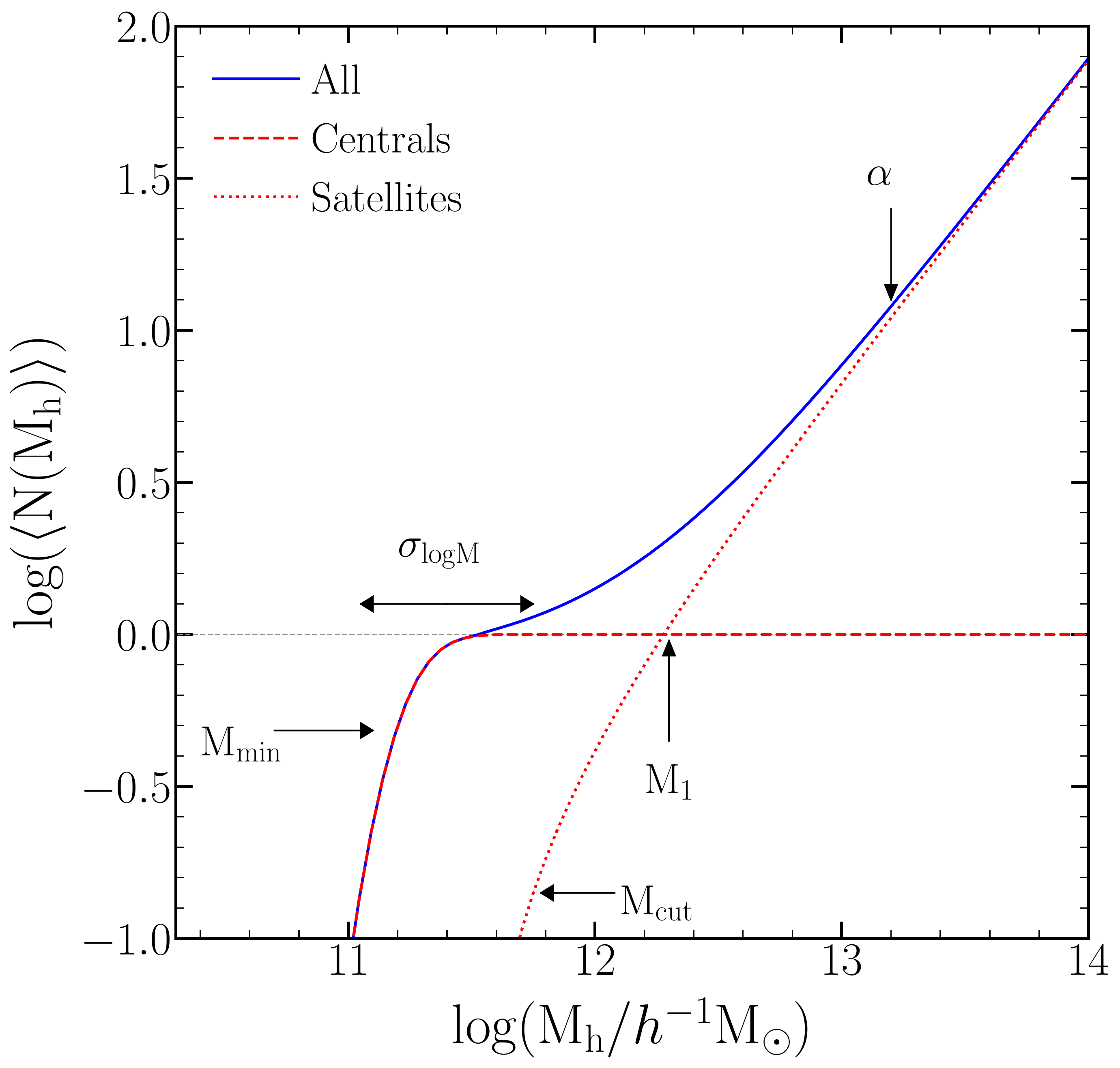}
\caption{ A schematic depiction of the standard 5-parameter form of the halo occupation function, which gives the mean number of galaxies per halo as a function of the host halo mass (based on Fig.~1 of \citetalias{C17}). The solid blue line represents the occupation function for all galaxies, which can be further separated into the contributions of central galaxies (red dashed line) and satellite galaxies (red dotted line). As a reference, we show an abundance of $\langle N_{\rm gal}(M_{\rm h})\rangle = 1$ as a thin horizontal grey dashed line; this will be shown in all subsequent HOD plots. The halo occupation function of central galaxies exhibits a gradual transition from zero to one galaxy per halo, which is well described by two parameters: $\Mmin$, the halo mass at which half of the haloes are populated by a central galaxy, and $\sigmaLogM$, which characterises the smoothness of this transition. The satellites occupation function is described by a transition from zero galaxies to a power law with three parameters, $\Mcut$, the minimum halo mass for hosting satellites; $\Mone$, the mass at which there is, on average, one satellite galaxy per halo; and the power-law slope $\alpha$. See text for more details and the explicit equations. }
\label{Fig:HODexample}
\end{figure}

\subsubsection{HOD parameterization} 
\label{SubSubSec:Param}

To parameterize the HOD, it is useful to distinguish between central galaxies, i.e. the primary galaxy at the centre of the halo, and the additional satellite galaxies, and to consider the contributions of each separately \citep{Kravtsov:2004, Zheng:2005}.
By definition, a dark matter halo cannot contain more than one central galaxy, but there is no upper limit on the number of satellites. Furthermore, for samples defined by properties that scale with halo mass, such as stellar mass or luminosity, a halo is typically populated first by a central galaxy, followed by additional satellite galaxies (although there can be haloes populated by only satellite galaxies in a given sample; see e.g., \citealt{Jimenez:2019, Chaves-Montero:2023}).

The traditional shape for the HOD is a smooth transition from zero to one galaxy for the central galaxies and a transition from zero to a power law for the satellites. The 5-parameter model introduced by \citet{Zheng:2005} (see also \citealt{Zheng:2007, Zehavi:2011}) is one of the most widely used parameterizations as it describes well samples of galaxies brighter than a given luminosity or more massive than a given stellar mass. We use this form of the halo occupation function in this work to describe the TNG300 and the SHAM mocks.
 
The mean occupation function of the central galaxies is described as a step-like function with a softened cutoff profile, to account for the dispersion between the stellar mass (or luminosity) and halo mass. It has the following form:
\begin{equation}
 \langle N_{\rm cen}(M_{\rm h})\rangle = \frac{1}{2}\left[ 1 + {\rm erf} \left( \frac{\log M_{\rm h} - \log M_{\rm min}}{\sigma_{\log M}}  \right) \right],
\label{Eq:Cen_HOD}
\end{equation}
where $M_{\rm h}$ is the host halo mass and 
$ {\rm erf}(x)$ is the error function,
\begin{equation}
 {\rm erf}(x) = \frac{2}{\sqrt{\pi}} \int_{0}^{x} e^{-t^2} {\rm d}t.
\end{equation}
The parameter $\Mmin$ characterizes the minimum mass for hosting a central galaxy above a given threshold, or more specifically, the halo mass at which half the haloes are occupied by a central galaxy (i.e., $\langle N_{\rm cen}(M_{\rm min})\rangle = 0.5$). The second parameter $\sigmaLogM$ represents the ``sharpness'' (width) of the transition from zero to one galaxy per halo. The value of $\sigmaLogM$ indicates the amount of scatter between stellar mass and halo mass.

The occupation function for satellite galaxies is modelled as:
\begin{equation}
 \langle N_{\rm sat}(M_{\rm h})\rangle = \left( \frac{M_{\rm h}-M_{\rm cut}}{M^*_1}\right)^\alpha,
\label{Eq:Sat_HOD}
\end{equation}
with $M_{\rm h}>\Mcut$, representing a power-law shape with a smooth cutoff at low halo masses. In this equation, $\alpha$ is the slope of the power law, which is typically close to one, $\Mcut$ is the satellite cutoff mass scale (i.e., the minimum mass of haloes hosting satellites), and $M^*_1$ is the normalisation of the power law.  A related parameter, $\Mone = \Mcut + M^*_1$, is often used to characterise the halo mass scale for hosting satellite galaxies above a given threshold. 
Specifically, it measures the average halo mass for hosting one satellite galaxy (i.e., $\langle N_{\rm sat}(M_{1})\rangle = 1$). In what follows, for clarity, we show results for $\Mone$. Nonetheless $M^*_1 >> \Mcut$, so $\Mone \sim M^*_1$. We have verified that all trends identified for $\Mone$ also hold for $M^*_1$.

The occupation functions for central and satellite galaxies can be fitted 
separately with these definitions (i.e. equations 1 \& 3), since the two occupation functions are independent of each other in terms of the parameters involved, with the total number of galaxies given by the sum of the central and satellite occupations: 
\begin{equation}
 \langle N_{\rm gal}(M_{\rm h})\rangle =  \langle N_{\rm cen}(M_{\rm h})\rangle +  \langle N_{\rm sat}(M_{\rm h})\rangle.
\end{equation}
Figure~\ref{Fig:HODexample} depicts a schematic representation of the shape of the HOD illustrating which features are sensitive to these five parameters.

We note that often a variant of these expressions is used, such that the cutoff profile for the central galaxies occupation is applied also to the satellite occupation, assuming (using our notation) that the total number of galaxies is given by $\langle N_{\rm cen}\rangle(1+\langle N_{\rm sat}\rangle)$. In that case, the fitting of the HOD cannot be done separately for centrals and satellites (because of the $\langle N_{\rm cen} \rangle \langle N_{\rm sat} \rangle$ term). Hence, assuming this form results in a more complex procedure to determine the best-fitting values of the parameters and ultimately gives poorer constraints, particularly for $\Mcut$. Furthermore, \citet{Jimenez:2019} show that satellite galaxies from a stellar mass-selected sample sometimes populate haloes whose central galaxies are not part of that sample. Assuming this form {(i.e. $\langle N_{\rm cen}\rangle(1+\langle N_{\rm sat}\rangle)$)} doesn't allow to account for such cases, and thus might bias the results. For these reasons, we choose to proceed with the formalism as detailed above in Eq.~2-4. We caution that one must be careful when comparing results obtained with different definitions.

\begin{figure}
\includegraphics[width=0.45\textwidth]{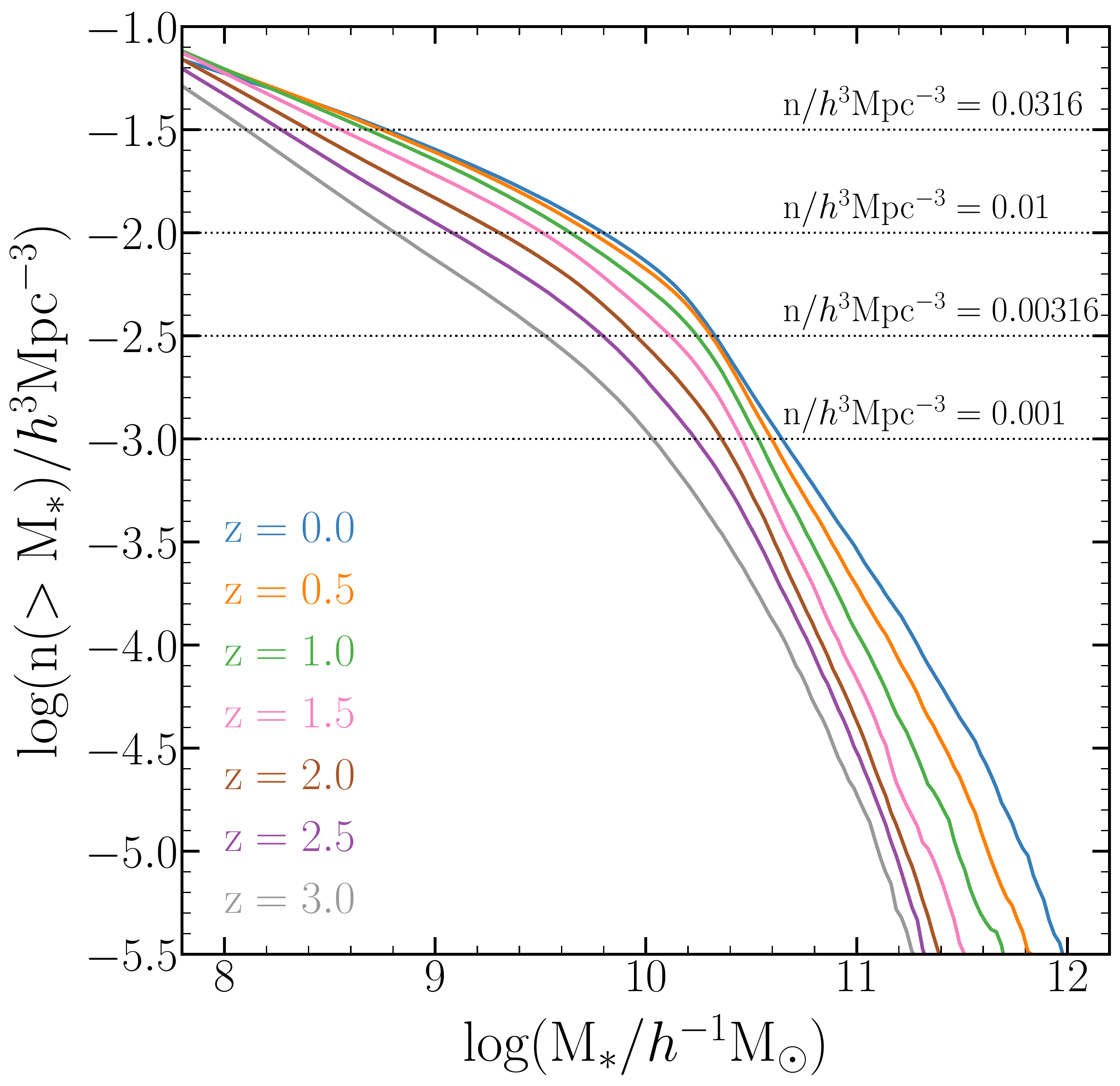}
\caption{ Cumulative stellar mass functions for galaxies in the TNG300 simulation. The coloured lines represent different redshifts as labelled. The dashed horizontal lines denote the number density samples adopted in this work (the values are marked at the upper right of each line). The galaxies selected for a given number density and redshift are those to the right of the intersection with their associated dashed line.}

\label{Fig:SMF}
\end{figure}

\begin{figure}
\includegraphics[width=0.45\textwidth]{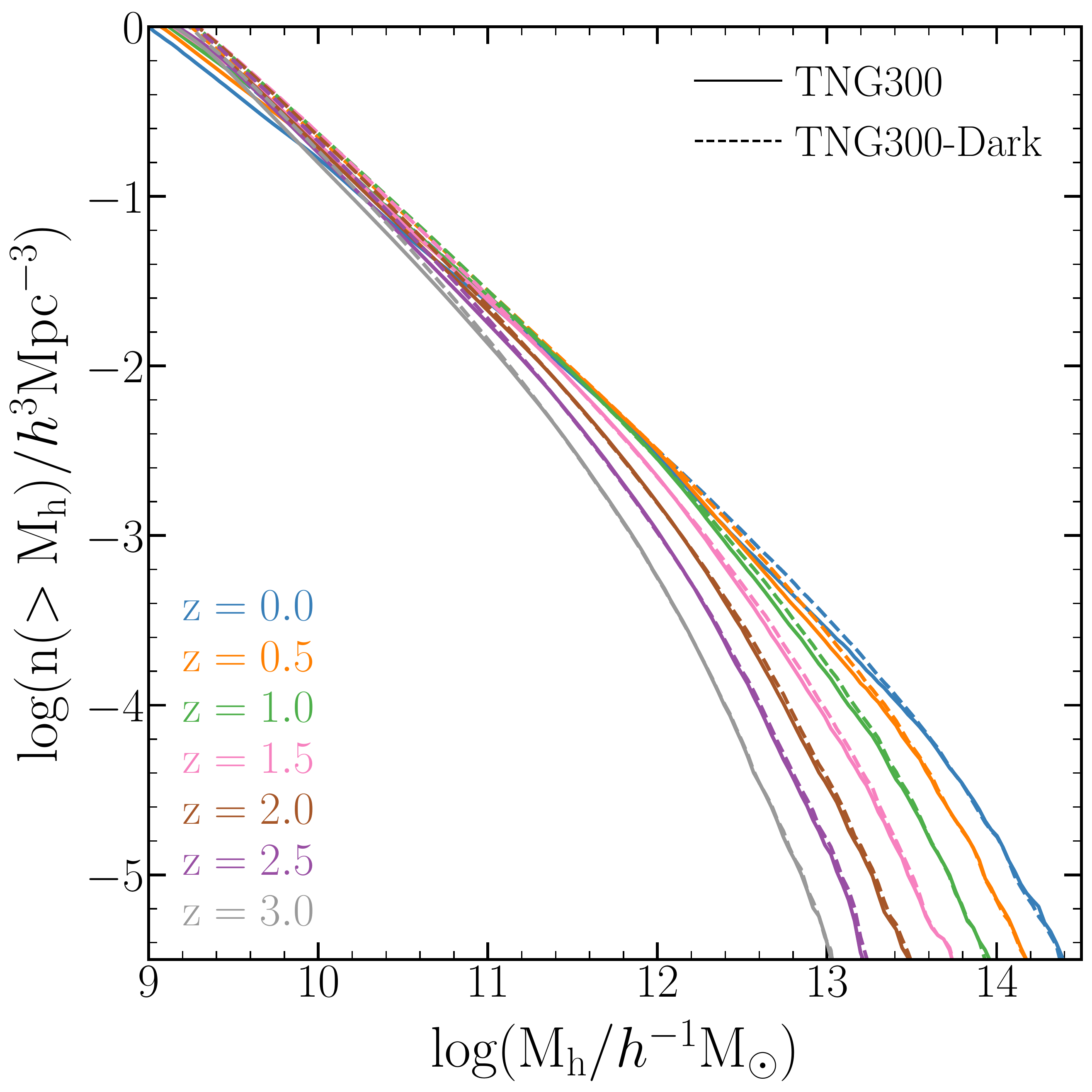}
\caption{ Cumulative halo mass functions for TNG300 (solid lines) and TNG300-Dark (dashed lines). The coloured lines correspond to different redshifts as labelled. Halo mass is defined as the total mass enclosed in a sphere whose mean density is 200 times the critical density of the Universe (also known as $\rm M_{200,crit}$). To compare the two samples, we calibrate the halo masses of the TNG300 by matching the halo mass functions (see \S~\ref{subsec:nden} for more details).}
\label{Fig:HMF}
\end{figure}

\subsection{Galaxy samples}
\label{subsec:nden}
We study stellar-mass selected galaxy samples corresponding to four different number densities and seven different redshifts. To construct these samples, we choose, at each epoch, the most massive galaxies corresponding to the following number densities: $0.0316$, $0.01$, $0.00316$ and $0.001$ $\ihMpcC$.  At $z=0$, these correspond to stellar mass thresholds of $6.05\times10^{8}$,  $6.47\times10^{9}$,  $2.19\times10^{10}$ and $4.54\times10^{10}$ $\hMsun$, respectively. The stellar mass of a galaxy in the TNG300 is defined as the sum of the masses of all stellar particles within twice the half-mass radius. We remind the reader that, since we are using the same stellar mass function for both the hydrodynamical and SHAM models, they will share the same cut for each number density. 

Fig.~\ref{Fig:SMF} shows the cumulative stellar mass functions for the 7 redshifts used in this work, $z=0,\ 0.5,\ 1.0,\ 1.5,\ 2.0,\ 2.5\ \&\ 3.0$. The horizontal dashed lines correspond to the four different number densities. The galaxies included in each sample are the ones to the right (more massive) of the intersection of these horizontal lines and the cumulative stellar mass functions. We chose these cuts to facilitate the comparison with \citetalias{C17}, where we analyzed galaxies from semi-analytical models selected in a similar fashion.

In order to facilitate the comparison of the HOD evolution for the different models, it is also necessary to correct the halo mass function of TNG300. Due to baryonic effects, The TNG300's halo mass function is not identical to that of TNG300-Dark, on which the SHAM mock was run. The cumulative halo mass functions for these two simulations are shown in Fig.~\ref{Fig:HMF}, for the different redshifts. To facilitate the comparison, we recalibrate the halo mass function of TNG300 to match that of the dark matter-only simulation. This is done by measuring the difference in halo mass between the simulations for each cumulative abundance, and then applying this difference to the TNG300's haloes. This step is particularly helpful for interpreting the evolution of the HOD parameters that represent masses (such as $\Mmin$, $\Mone$, and $\Mcut$), given that the differences between the halo mass functions are not constant with redshift. All TNG300 results presented in this paper incorporate this correction.

\begin{figure}
\includegraphics[width=0.43\textwidth]{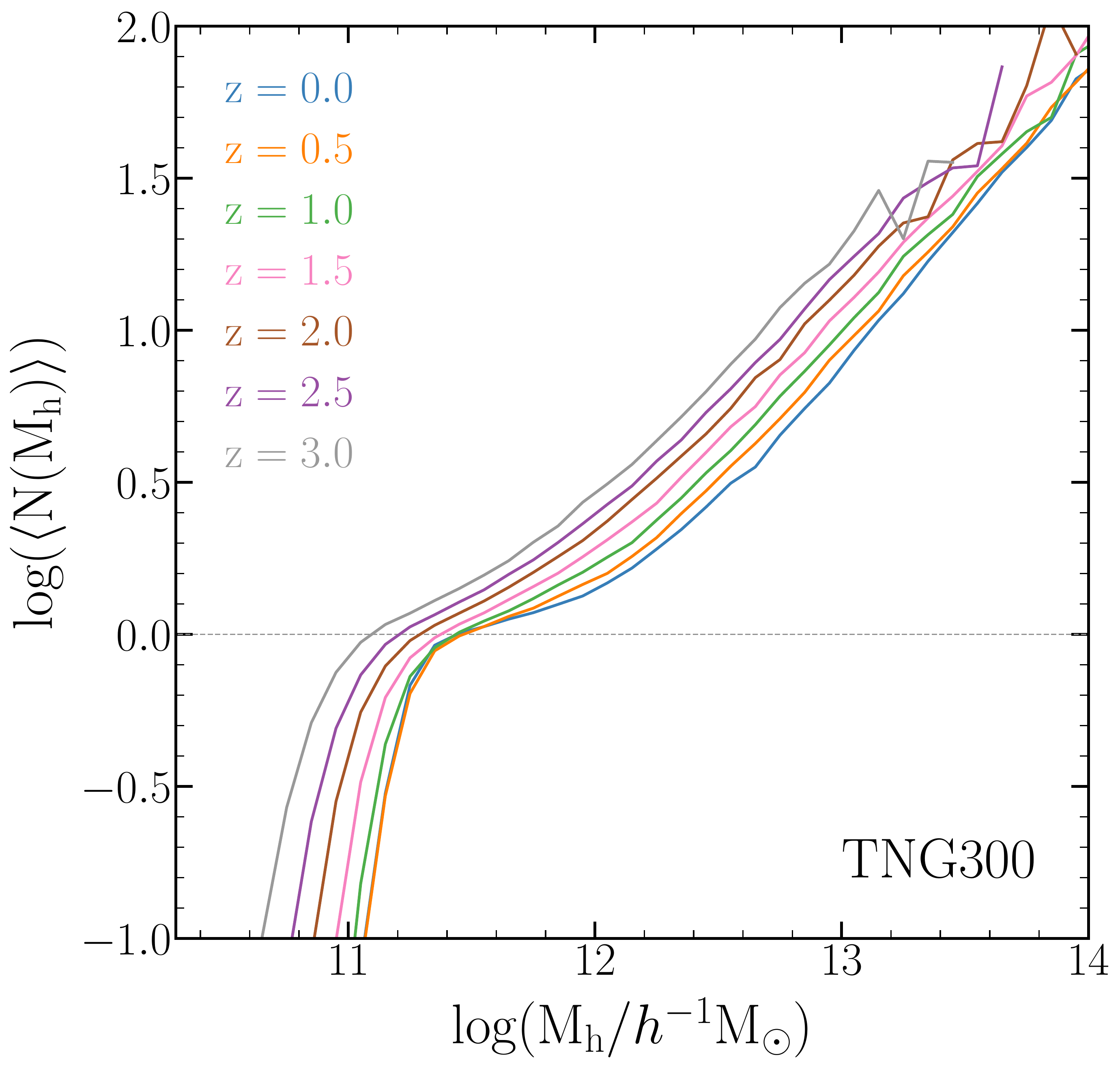}
\includegraphics[width=0.43\textwidth]{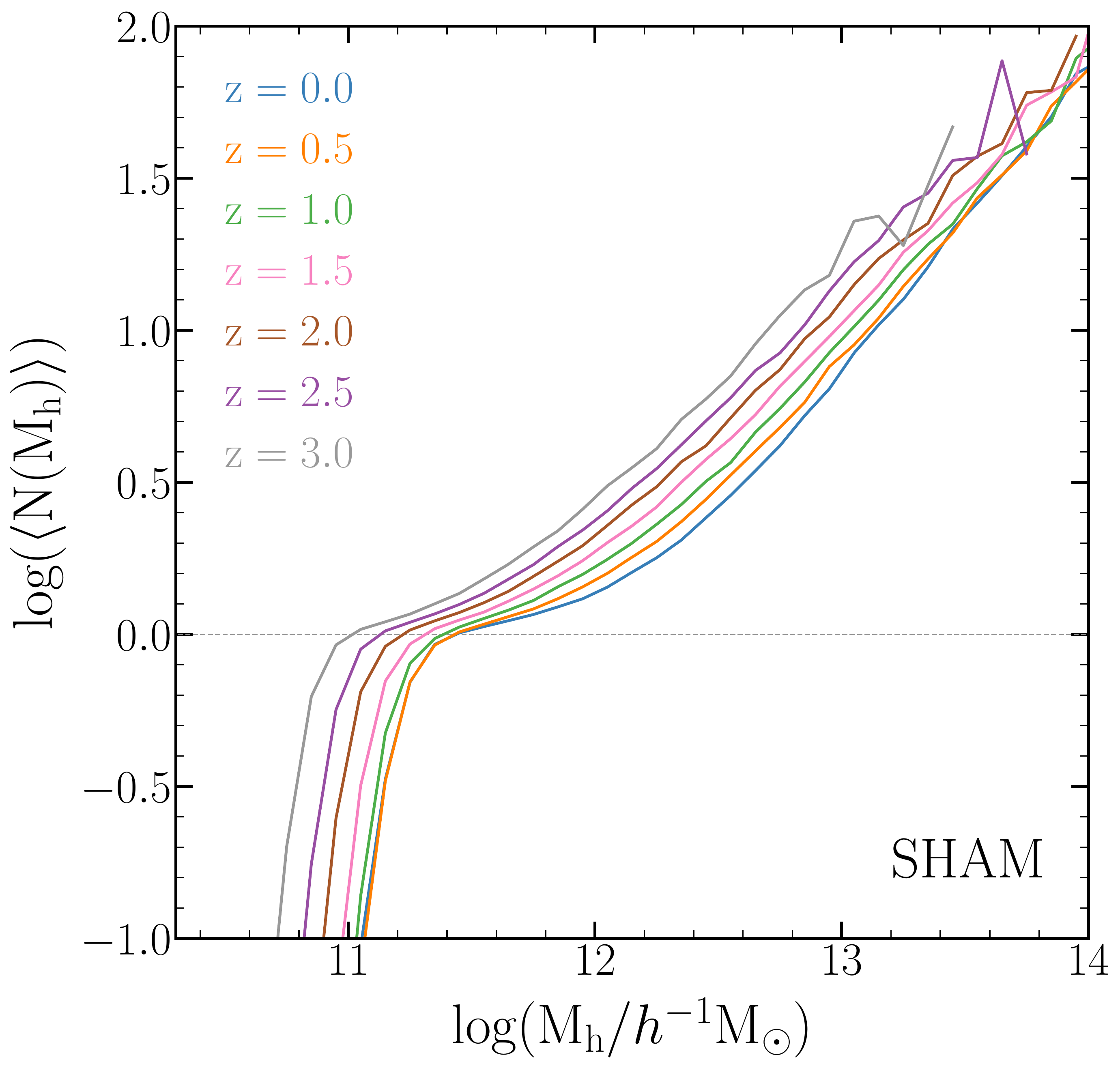}
\includegraphics[width=0.43\textwidth]{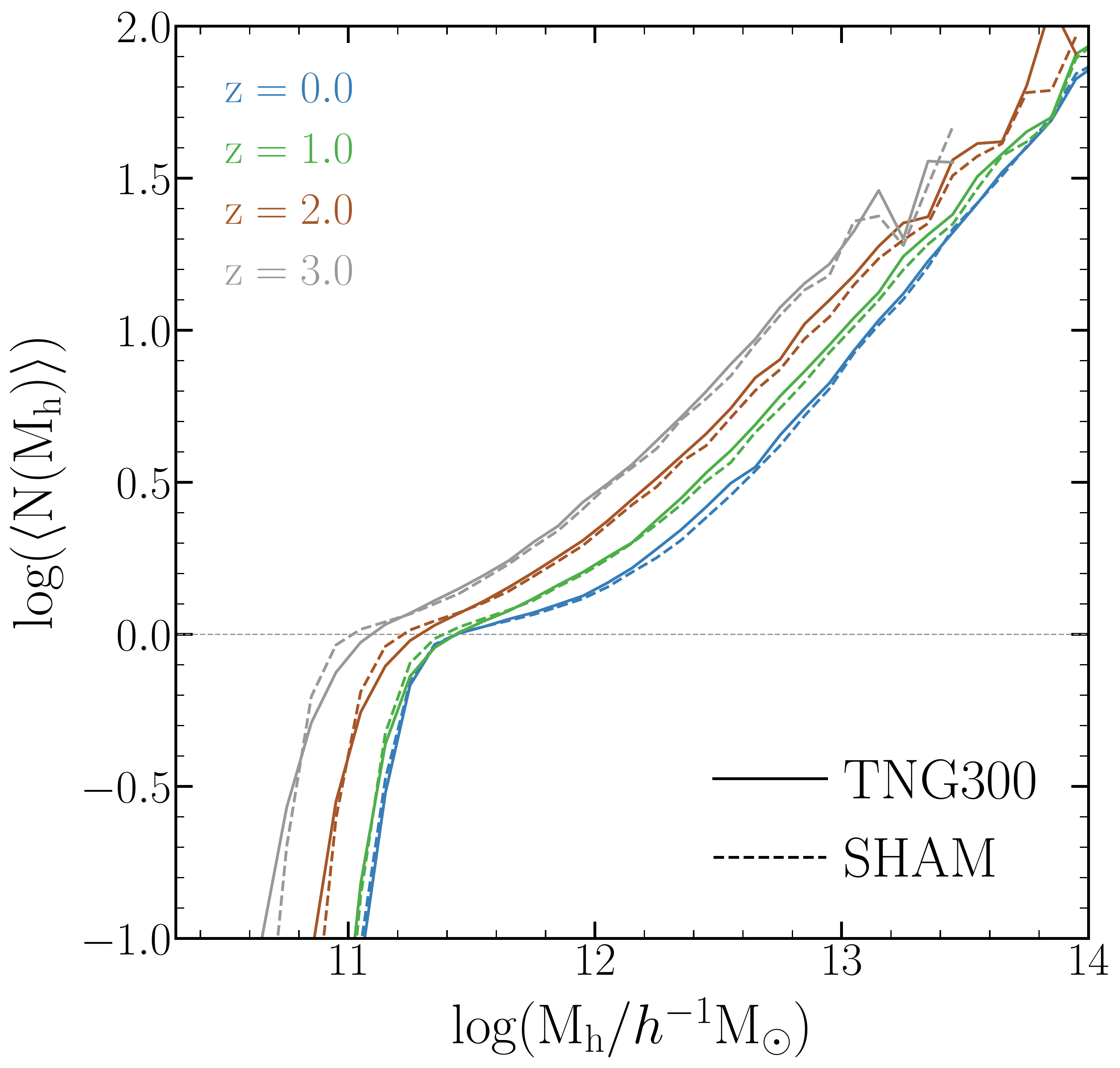}
\caption{The HODs in the TNG300 simulation (top panel) and for a mock galaxy sample run with a SHAM model (middle panel), for stellar-mass selected galaxies corresponding to a number density of $\rm 0.0316\ h^3Mpc^{-3}$. The different lines correspond to different redshifts spanning $\rm z =0$ to $3$, as labelled.  To facilitate the comparison between the models, we show in the bottom panel the HODs for both the TNG300 and the SHAM mock at $\rm z=0,\ 1,\ 2$ and $3$.}
\label{Fig:HOD_sample}
\end{figure} 

\begin{figure} 
\includegraphics[width=0.45\textwidth]{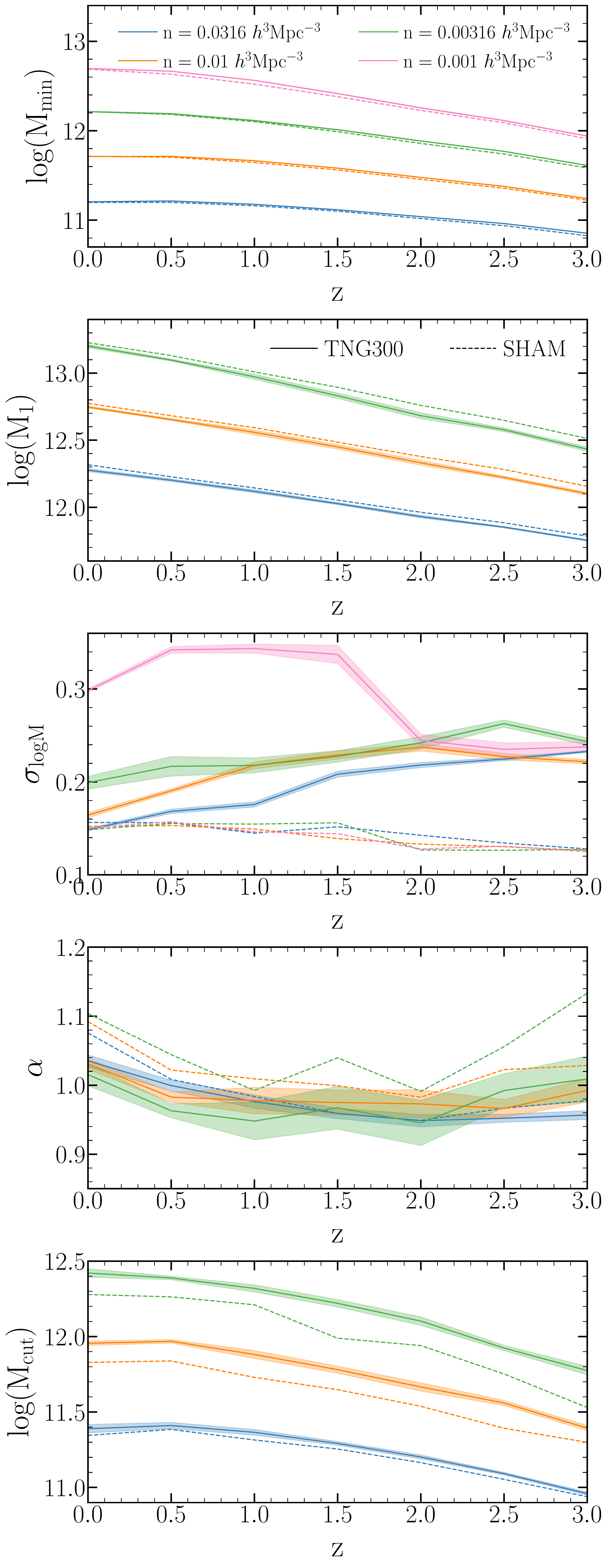}
\caption{Redshift evolution of the 5 fitting parameters of the HOD, corresponding to the TNG300 simulation (solid lines) and the SHAM mock (dashed lines). From top to bottom, the properties shown in each panel are $\Mmin$, $\Mone$, $\sigmaLogM$, $\alpha$ and $\Mcut$. Different colours represent different number density samples, as labelled. For the lowest number density, $n=0.001\ \ihMpcC$, we only show the parameters corresponding to the centrals occupation ($\Mmin$ and $\sigmaLogM$), since the constraints on the satellites occupation parameters are poor at higher redshifts (due to the limited amount of haloes with satellite galaxies). Error bars represent the standard deviation from the fitted parameter value.}
\label{Fig:Fit_ev} 
\end{figure}

\section{The evolution of the HOD}
\label{sec:HODev}
We compute the halo occupation functions in the TNG300 simulation and for the SHAM model for the four number density samples previously stated and at the seven redshift outputs between z=0 and z=3. Please note that we are here directly measuring the average halo occupation as a function of halo mass, as opposed to inferring it from the correlation function, as is typical in galaxy clustering studies. 

In Fig.~\ref{Fig:HOD_sample}, we show the HODs for the galaxy samples with a number density of $n=0.0316\ \ihMpcC$ at the seven redshift outputs between $z=0$ and $z=3$. The top and middle panels show the HODs for the TNG300 and the SHAM model, respectively, while the bottom panel compares the HODs of both models for a smaller set of redshifts. The evolution of the HOD in both models appears similar. The overall trend is a shift of the halo occupation function toward larger halo masses with decreasing redshift (increase in time).  
In more detail, for both models, the threshold for hosting central galaxies (at the lower occupation region at low halo masses), increases monotonically with time from $z=3$ to $z\sim1$. From that point until $z=0$, the shift in halo mass diminishes with the threshold for starting to host galaxies remaining similar. In contrast, the satellite occupation appears to continuously shift with decreasing redshift. These results are consistent with the findings of \citetalias{C17} in semi-analytic galaxy formation models.

To gain a better understanding of the evolution of HODs, we fit the halo occupation functions using the 5-parameter functional form described in  \S~\ref{SubSubSec:Param} and analyse the evolution of those parameters. We fit the central and satellite occupations independently, and assume a constant error per bin for the fitting. In previous works (e.g., \citealt{C13}, \citetalias{C17}), we tested using different weights on each of the points of the HOD (such as weighting by the abundance of haloes or the effective bias in each mass bin), finding that this tends to overemphasize a specific part of the HOD, resulting in significant discrepancies at high halo masses. We estimate the error on the HOD fitting parameters by normalizing the constant bin errors such that the best fit has a reduced chi-square of one (i.e., $\chi^2_{\rm min}/{\rm d.o.f} = 1$). 

Fig.~\ref{Fig:Fit_ev} presents the values for the best fitting parameter of the HOD, $\Mmin$, $\Mone$, $\sigmaLogM$, $\alpha$ and $\Mcut$, as a function of redshift. The solid lines show the values for the TNG300 while the SHAM predictions are shown as dashed lines. The different colours represent different number densities as labelled. We do not show the prediction for the lowest number density for the satellite HOD parameters, since the number of haloes with satellite galaxies at high redshift was too low to do a proper fit.

While there are some differences between the values of the parameters for the TNG300 and the SHAM at different redshifts, overall there is a good agreement between the models for all but one parameter, $\sigmaLogM$. While the SHAM technique is known for being able to reproduce the galaxy clustering (and therefore, the HOD) of complex galaxy samples as a hydrodynamic simulation (e.g., \citealt{ChavesMontero:2016, C21c}), it is surprising that even in its most basic form (without scatter), the model is in good agreement with the TNG300 predictions. We remind the reader that the SHAM model does not incorporate any baryonic processes and that the properties of the resulting galaxy population depend solely on the gravitational growth of the subhaloes in the simulation. This in turn depends on the cosmological model corresponding to the dark matter-only simulation used.

More significant than the overall agreement of the values of the HOD parameters for these two different models, is the common way these four parameters evolve which we summarize as follows:
\begin{itemize}
\item  The characteristic halo mass for hosting a central galaxy, $\Mmin$, increases linearly (in logarithmic scale) from $z=3$ to $z\sim 1-0.5$ and then remains constant until $z=0$.
\item The characteristic halo mass for hosting a satellite galaxy, $\Mone$, increases linearly (in logarithmic scale) from $z=3$ to $z=0$.
\item The power-law slope of the satellites occupation, $\alpha$, is roughly constant from  $z=3$ to $z\sim 1-0.5$, and then increases linearly until $z=0$. There are some differences in the behaviour of $\alpha$ in the TNG300 and the SHAM, which are however not that significant given the level of noise in this parameter.  
\item The satellites cutoff mass scale, namely the minimum mass of haloes hosting satellites, $\Mcut$, increases linearly (in logarithmic scale) from $z=3$ to $z\sim 1-0.5$, and then stays constant until $z=0$ (the same as $\Mmin$).
\end{itemize}

These common evolutionary trends exhibited by the TNG300 and the basic SHAM, which also coincide with those found by \citetalias{C17} for two different SAMs, are the most important results of this work. From this, we can conclude that these evolutionary trends are independent of galaxy formation physics, meaning that they can be used to populate HODs in simulated lightcones regardless of galaxy formation physics. We emphasise that we only claim that the functional form for the evolution of these parameters is universal. We do not maintain that either the value of the HOD parameters or the amplitude of their evolution with redshift (i.e., the rate at which these parameters change with redshift) are universal. Typically, these parameters are fit from galaxy clustering, and their values can vary between galaxy samples. Still, given the large similarities between two distinct models, such as the TNG300 and a basic SHAM, the way the HOD parameters evolve cannot be regarded as a good proxy for galaxy formation physics.

To further assess the robustness of our results, we also examine the evolution of these parameters in the EAGLE hydrodynamical simulation \citep{Schaye:2015,Crain:2015}, as presented in Appendix~\ref{sec:EAGLE}. EAGLE has a smaller volume but a higher resolution than the TNG300, and it was executed with an SPH code rather than an adaptive mesh code like the one used in the TNG300. We find similar evolutionary trends as the ones observed for the TNG300, the SHAM model, and the SAMs. 
We have additionally studied the evolution of the HOD in TNG300 when selecting galaxies by $r$-band magnitude instead of stellar mass, finding again similar evolutionary trends. We refer the reader to \S~3.4 of \citetalias{C17} for a more in-depth analysis of the evolution of these parameters and a simple parameterization of the evolution of the HOD parameters that can be used in the construction of galaxy samples or the interpretation of clustering data at different redshifts.

As for $\sigmaLogM$, this property measures the scatter between the halo mass and stellar mass of a galaxy sample. The prediction of a SHAM without scatter will only measure the dispersion between the halo mass and $\vpeak$, which is significantly smaller than the expected scatter between stellar mass and halo mass of a fully physical galaxy formation model. As concluded from previous work (e.g., \citetalias{C17}), this parameter should be the one that best captures the physics of galaxy formation of a galaxy sample. However, as demonstrated in \citet[][see also \citealt{Tinker:2011, Parejko:2013, C23a}]{Zehavi:2011}, this parameter, along with $\Mcut$, have the weakest constraint from galaxy clustering. There are two reasons why this parameter does not significantly impact the
galaxy clustering signal: (a) For a fixed value of $\Mmin$, larger values of $\sigmaLogM$ increase the number of less massive haloes occupied and decrease the number of massive haloes occupied. For reasonable values of $\sigmaLogM$, the change in the halo occupation will not produce a significant change in the bias of the sample, and thus, will not impact the galaxy clustering. (b) While it is the HOD parameter that captures most of the galaxy formation physics of a model, it does not capture the intrinsic dependence of the stellar mass on the assembly history of the haloes, for fixed values of halo mass. This dependence has been incorporated in some extensions of the HOD model (e.g., \citealt{Hearin:2016, Xu:2020}) which can help to improve the galaxy clustering predictions of a HOD but are not normally included in HOD models for lightcones. 

Because of the poor constraint on $\sigmaLogM$, the lack of agreement of its evolution among different galaxy formation models, and its low impact on galaxy clustering, it is not required to model $\sigmaLogM$ perfectly when creating HOD-based mock catalogues. Nonetheless, the values appear relatively constant with redshift, which makes sense given that we do not anticipate a significant change in the evolution of the stellar mass-halo mass relationship. This is one of the foundations of the SHAM model (see \citealt{C15} for further discussion). 

 In \citetalias{C17}, based on the evolution of the HOD parameters found in the SAMs, we proposed an evolution model for each of the parameters of the HOD, using their values at $z=0$ and up to one additional evolution parameter. In that work, we have tested this parametrization, finding that we can accurately recover the HOD at $z=1$, evolved from $z=0$, for several number densities.

\begin{figure}
\includegraphics[width=0.45\textwidth]{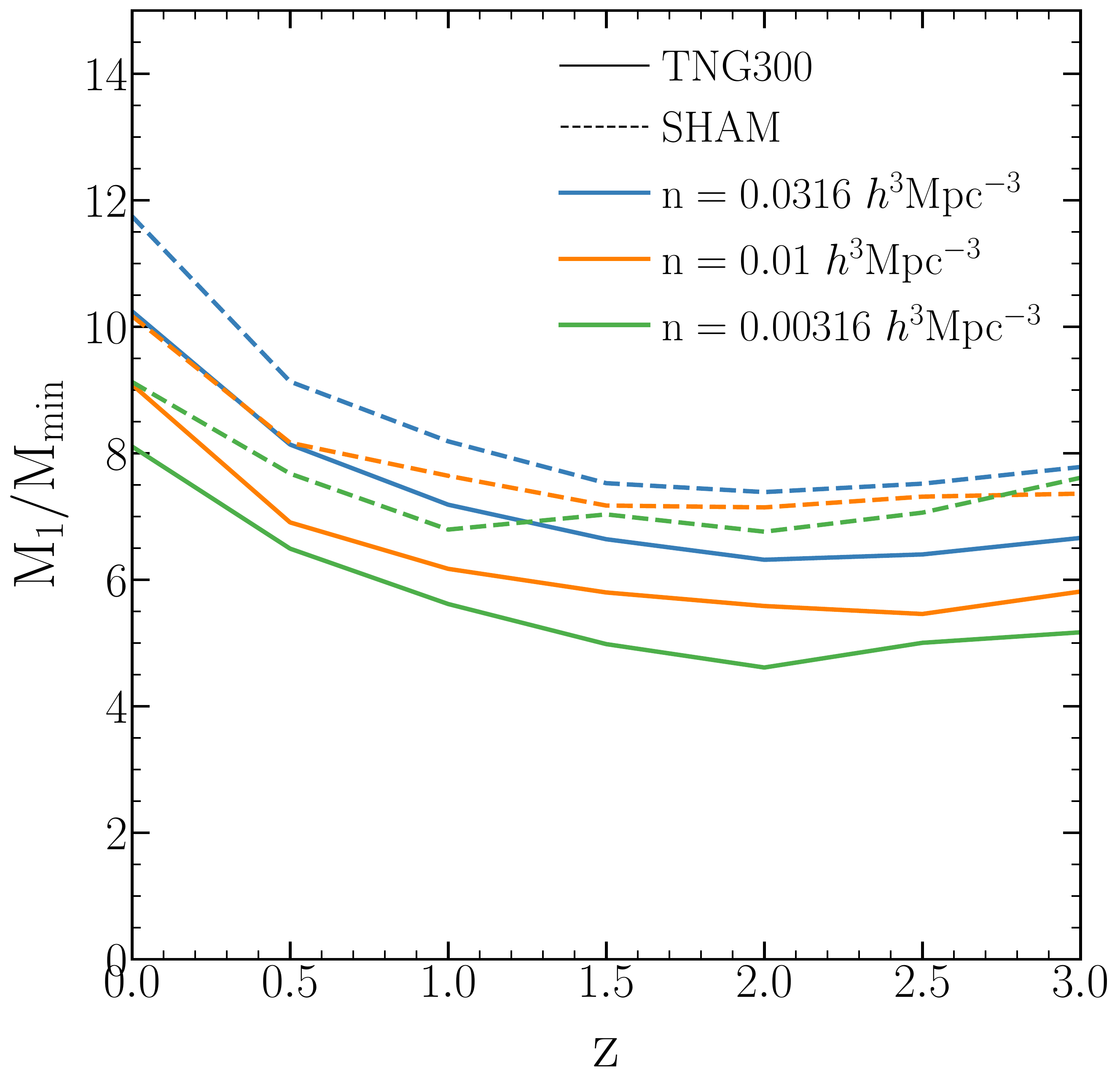}

\caption{ Redshift evolution of the ratio of the two characteristic halo mass parameters of the HOD, $\Mone$ and $\Mmin$. The predictions from the TNG300 simulation are shown as solid lines, while the dashed lines denote the results from the SHAM mock. The different colours represent different number densities as labelled. }
\label{Fig:M1_Mmin}
\end{figure}

\section{Origin of the HOD evolution}
\label{sec:discussion}

In \S~\ref{sec:HODev} we studied the evolution of the HOD in the TNG300 hydrodynamical simulation and in a SHAM applied to the dark matter-only simulation, finding that the evolution of the HOD parameters is largely the same.  Since no galaxy formation physics is included in our SHAM implementation and it lacks any free parameter that attempts to reproduce the impact of baryonic physics (such as a scatter in the $\vpeak$-stellar mass relation, modifying the dynamical friction model, etc.), it appears that the evolution is independent of galaxy formation physics. This is further corroborated by the overall agreement with results from the EAGLE hydrodynamical simulation (Appendix~\ref{sec:EAGLE}) and SAMs applied to the Millennium Simulation (\citetalias{C17}). This leads us to conclude that the evolution of the HOD is instead dominated by the cosmological model and the hierarchical growth of structure.

We would still like to discern which aspect of the cosmological picture shapes the evolution of the HOD parameters. One possibility is that, at least for the parameters that represent halo masses (such as $\Mmin$, $\Mone$, and $\Mcut$), the evolution arises from the typical growth of haloes. To determine this, we examined the evolution of these parameters as peak height values ($\nu(M,z)=\delta(M)/\delta(z)$) rather than halo masses (not shown here). If, in theory, the variation of these parameters with redshift ends up being smaller or even null when expressed by their peak height, using HODs as a function of peak height instead of mass will facilitate their implementation on lightcones mocks. However, we found that the changes with redshift are stronger when expressed in terms of peak height. This is because the evolution of the halo mass function is stronger than the evolution of the HOD parameters. We, therefore, conclude that it is not practical to express these parameters in terms of peak height.

Another factor that can potentially influence how HODs evolve is the values of the cosmological parameters. This is a plausible explanation of the agreement since the TNG300-Dark simulation (on which we run the SHAM mock) and the TNG300 simulation share the same cosmology. Moreover, the growth of dark matter haloes is affected by cosmology. A strong cosmological dependence of the evolution of HODs with any cosmological parameter could imply that we could constrain cosmology based on the HOD evolution we infer from galaxy surveys. However, when examining the evolution of the HOD in SHAM mocks built on simulations with different cosmologies, we find only small changes in the evolution of the parameters, and, more importantly, that they all share the same functional form as the one found for our fiducial cosmology. The details of this analysis are presented in Appendix~\ref{sec:cosmo_HODparam} for eight cosmological parameters. This indicates that the specific flavour of the cosmological model within the $\Lambda$CDM framework also does not influence much the evolution of the HOD.

Since the details of the cosmological model do not have a significant impact on how the HOD evolves, we deduce that this evolution is governed by the hierarchical way in which haloes and subhaloes (and therefore galaxies) form and evolve in the $\rm  \Lambda CDM$ model. This becomes more apparent when we examine the evolution of the ratio of the two halo mass parameters $\Mone$ and $\Mmin$, which is frequently employed to characterise a galaxy population (e.g., \citealt{Zehavi:2011, Coupon:2012, Guo:2014, Skibba:2015}). This ratio represents the mass range over which a halo hosts only a central galaxy from the sample before hosting additional satellite galaxies, giving rise to the ``plateau'' in the halo occupation function (Fig.~\ref{Fig:HODexample}; see also \citealt{Watson:2011}).  Fig.~\ref{Fig:M1_Mmin} shows the evolution of this ratio for the TNG300 and our SHAM model, where the value of $\Mone/\Mmin$ roughly plateaus at high redshift and then increases with time, toward the present. 

\citetalias{C17} explored the change in this ratio when assuming alternative ``non-hierarchical'' evolution models for the galaxies.
The models they tested were a passive evolution model (e.g., \citealt{Seo:2008}), where galaxies no longer form or merge;
a tracking evolution, in which galaxies no longer form but continue to merge;
and a descendant clustering selection \citep{Padilla:2010} where galaxies are selected based on the evolution of halo clustering (see Fig.~11 in \citetalias{C17} and discussion thereof). 
All these models exhibit significantly different evolution for $\Mone/\Mmin$, with the ratio decreasing toward lower redshifts,  in contrast to the evolution found in our SHAM mocks and the TNG300.
The \citet{Guo:2013} SAM used in \citetalias{C17} also exhibits the same type of evolution as the models presented in this work. \footnote{We note that the \citet{Gonzalez-Perez:2014} SAM additionally used in \citetalias{C17} showed some variation in the evolution of $\Mone/\Mmin$. This is likely due to the different (sub)halo algorithms employed compared to \citet{Guo:2013}, TNG300 and TNG300-Dark, which use the {\tt FOF} and {\tt SUBFIND} algorithms (see \S~\ref{subsec:TNG300} for more details).} 
We would like to again emphasise that we only claim a common evolution of $\Mone/\Mmin$, not that all the models share the same values. Nonetheless, while this ratio could, in principle, contain information about the galaxy formation physics of a galaxy sample, the small differences found for the TNG300 and the SHAM model suggest it may not be the best proxy to characterise this information.

We conclude that the evolution of the HOD is independent of galaxy formation physics, or the specifics of the cosmological model. Any galaxy population that grows hierarchically, such as stellar mass selected galaxies, in a $\rm  \Lambda CDM$ (or similar) framework should exhibit similar evolutionary trends to the ones found in this work.

\section{Conclusion}
\label{sec:Summary}

In this paper, we look at the evolution of the HOD of stellar mass-selected galaxies from two different models; a magneto-hydrodynamic simulation, the TNG300, and a SHAM mock built from the dark matter-only simulation without any baryonic physics implemented.  We characterise the cosmic evolution by fitting the HODs at different redshifts with the standard 5-parameter parametric form \citep{Zheng:2005}. Our main findings are as follows: 

\begin{itemize}

\item The HODs for the TNG300 and the SHAM models are similar at all redshifts and number densities, exhibiting a similar evolution of the halo mass parameters. The one standout is $\sigmaLogM$, capturing the width of the transition from zero to one galaxy per halo, which varies between the models.
  
\item The values of $\sigmaLogM$ are different for the TNG300 and the SHAM model. This parameter is related to the scatter between halo mass and stellar mass and is expected to be dependent on the galaxy formation physics model. At the same time, this parameter has little effect on galaxy clustering \citep{Zehavi:2011} and thus it is not always essential to define its value or its evolution with high precision.

\item The evolution of the HOD is also similar to that measured in the EAGLE hydrodynamical simulation,  and for a ${\rm M_r}$ magnitude limited sample in the TNG300 simulation.  The evolution of the parameters (other than $\sigmaLogM$) is also similar to that of semi-analytical models of galaxy formation, as explored in \citetalias{C17}.

\item The evolution of the HOD is largely insensitive to variations of the cosmological parameters,  with only $\sigma_8$ and $w_0$ somewhat impacting the shape, and with a common evolutionary form for all parameters.
  
\item The values and evolution of the $\Mone/\Mmin$ ratio are similar for the TNG300 and the SHAM model. They are also in agreement with the ones found by \citetalias{C17}  when analysing a SAM with the same (sub)halo identification algorithm, but different to that found when assuming alternative galaxy evolution models (such as passive evolution).  

\end{itemize}

Based on these results, it appears that the physics of galaxy formation has little impact on the evolution of the HOD for stellar mass-selected samples. Given that the HOD and galaxy clustering of a SHAM model without scatter or any free parameter only depend on the cosmological model assumed in the dark matter simulation on which it is based, we can conclude that the cosmological framework dominates the HOD evolution for this type of galaxies. By cosmological framework here we specifically refer to the hierarchical building of haloes and galaxies, as we have also demonstrated that the values of the cosmological parameters have little impact on the HOD evolution.

The way the HOD parameters evolve in the SHAM model is a strong indication of how consistent and good the model is when populating galaxies at different redshifts, and the potential it has for creating mock galaxy catalogues (given sufficient resolution to follow the subhaloes). Furthermore, our results provide an important simplification to the process of creating mock galaxy catalogues over a large redshift range. They lend significant support for some of the ansatzes accepted today when generating mock galaxy catalogues on simulated lightcones, namely that the HOD evolution model is robust and needn't change based on the assumed galaxy formation model.
This robustness, in turn, can facilitate the HOD interpretation of clustering measurements at different redshifts from upcoming large galaxy surveys.

We clarify that the results presented here and subsequent conclusions have only been investigated for galaxy samples selected by stellar mass (and luminosity), that grow hierarchically. The HOD of galaxies selected, for example, by star formation rate may not follow the same pattern. We note that the extension of our work to that is not trivial, as it requires a somewhat more complex HOD model as well as a non-trivial extension of the SHAM methodology (S.\ Ortega Martinez, in prep.), and we reserve this for future work. 

\section*{Data Availability}

The IllustrisTNG simulations, including TNG300, are publicly available and accessible at \url{www.tng-project.org/data} \citep{Nelson:2019}. The data underlying this article will be shared on reasonable request to the corresponding author.

\section*{Acknowledgements}

We thank Nelson Padilla, Celeste Artale, Carlton Baugh, Peder Norberg, Shaun Cole and Alex Smith for useful comments and discussions.
We are grateful for the anonymous referee for their careful reading of the manuscript and insightful comments that helped improve the presentation and clarity of the work.
We acknowledge the hospitality of the ICC at Durham University and the helpful conversations with many of its members.
SC acknowledges the support of the ``Juan de la Cierva Incorporaci\'on'' fellowship (IJC2020-045705-I).
IZ was partially supported by a CWRU ACES+ Opportunity Grant.
The authors also acknowledge the computer resources at MareNostrum and the technical support provided by Barcelona Supercomputing Center (RES-AECT-2020-3-0014).

\bibliographystyle{mnras}
\bibliography{Biblio}

 

\appendix

\section{The evolution of the HOD in the EAGLE simulation} 
\label{sec:EAGLE}

\begin{figure}
\includegraphics[width=0.45\textwidth]{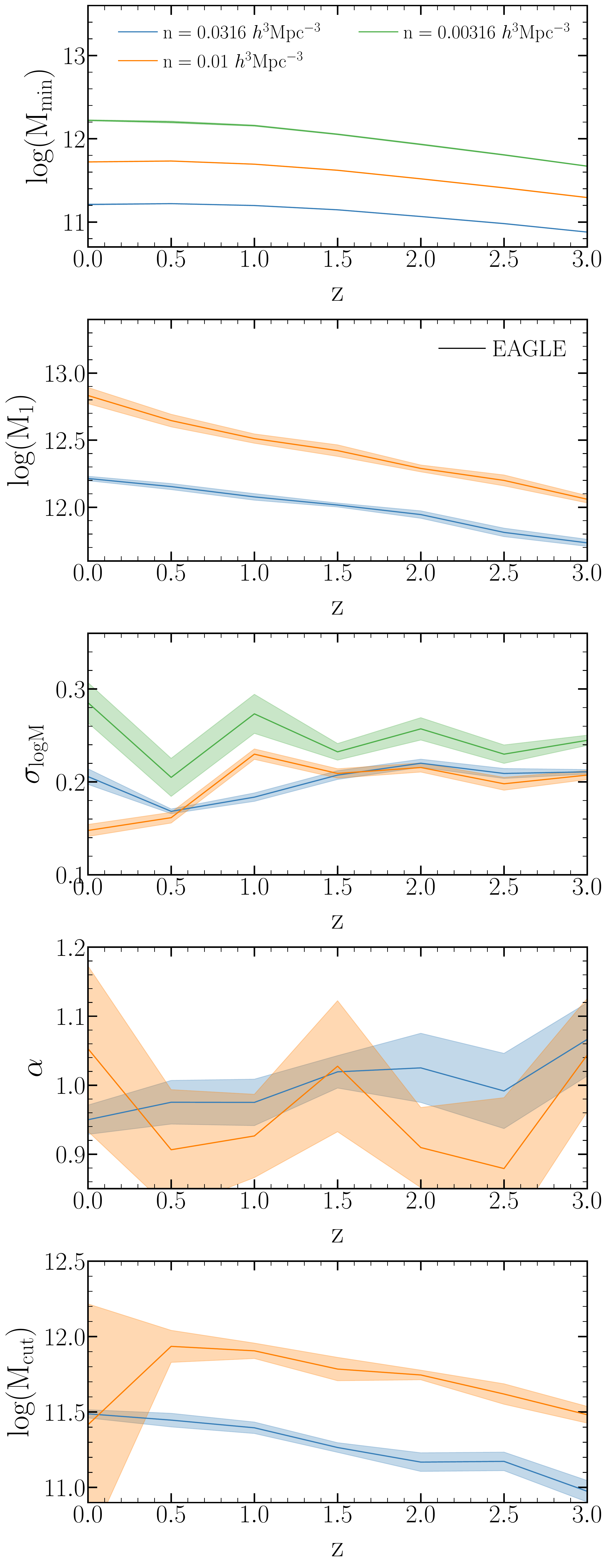}
\caption{The same as Fig~\ref{Fig:Fit_ev}, but now for the galaxies of the EAGLE hydrodynamical simulation. Only the three largest number densities are shown for the parameters of the central galaxies occupation and the two largest number densities for the parameters of the satellite occupation, due to the higher level of noise (smaller box).}
\label{Fig:EAGLE}
\end{figure}

In addition to the TNG300 hydrodynamic simulation, we also measure the evolution of the HOD in the EAGLE hydrodynamic simulation \citep{Schaye:2015, Crain:2015, McAlpine:2016}. This simulation was run using a modified version of the N-Body Tree-PM smoothed particle hydrodynamics (SPH) code {\tt GADGET 3} \citep{Springel:2005}, different from the adaptive mesh code \texttt{AREPO} \citep{AREPO} used for the TNG300. EAGLE has a smaller volume but better resolution than the TNG300 (2x1504 particles in $67.77\ \hMpc\equiv 100\ {\rm Mpc}$) and was calibrated to reproduce a different set of low-redshift observables.

Same as for the TNG300 and the SHAM mock, we fit the HOD at different redshifts and look at the evolution of the HOD fitting parameters. These results are presented in Fig~\ref{Fig:EAGLE}. Due to the limited volume of EAGLE, we only show the three largest number densities for the parameters that describe the occupation of central galaxies ($\Mmin$ and $\sigmaLogM$) and the two largest number densities for the parameters that describe the satellite occupation ($\Mone$, $\alpha$ and $\Mcut$). For simplicity, we did not shift the halo mass function to match its dark matter counterpart as we did with the TNG300. 

 find similar patterns for EAGLE, as those of the TNG300 and the SHAM mock. We clarify again that by the same patterns
we mean here that the redshift evolution of the HOD parameters follow the same trends as for the other galaxy formation models. Namely, $\Mmin$ and $\Mcut$ increase linearly (in logarithmic scale) from $z=3$ to $z\sim 1-0.5$ and then remain constant until $z=0$ and $\Mone$ increases linearly (in logarithmic scale) from $z=3$ to $z=0$. These findings support our hypothesis that the evolution of the HOD parameters is not significantly influenced by the physics of galaxy formation and that the use of an evolutionary formalism as the one proposed by \citetalias{C17} would be valid for most galaxy formation models.

\section{The evolution of the HOD for different cosmologies} 
\label{sec:cosmo_HODparam}

\begin{figure*}
\includegraphics[width=0.24\textwidth]{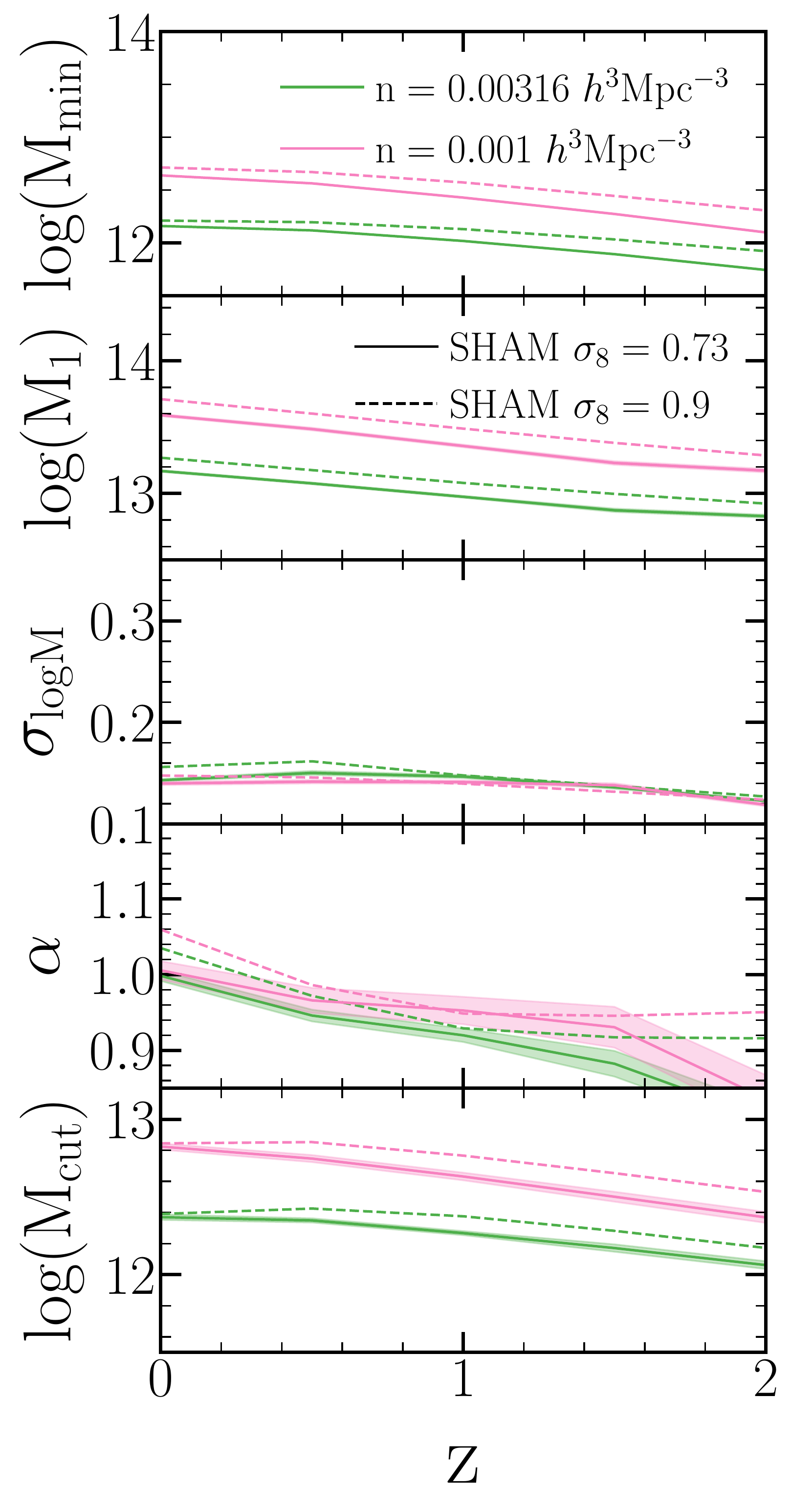}
\includegraphics[width=0.24\textwidth]{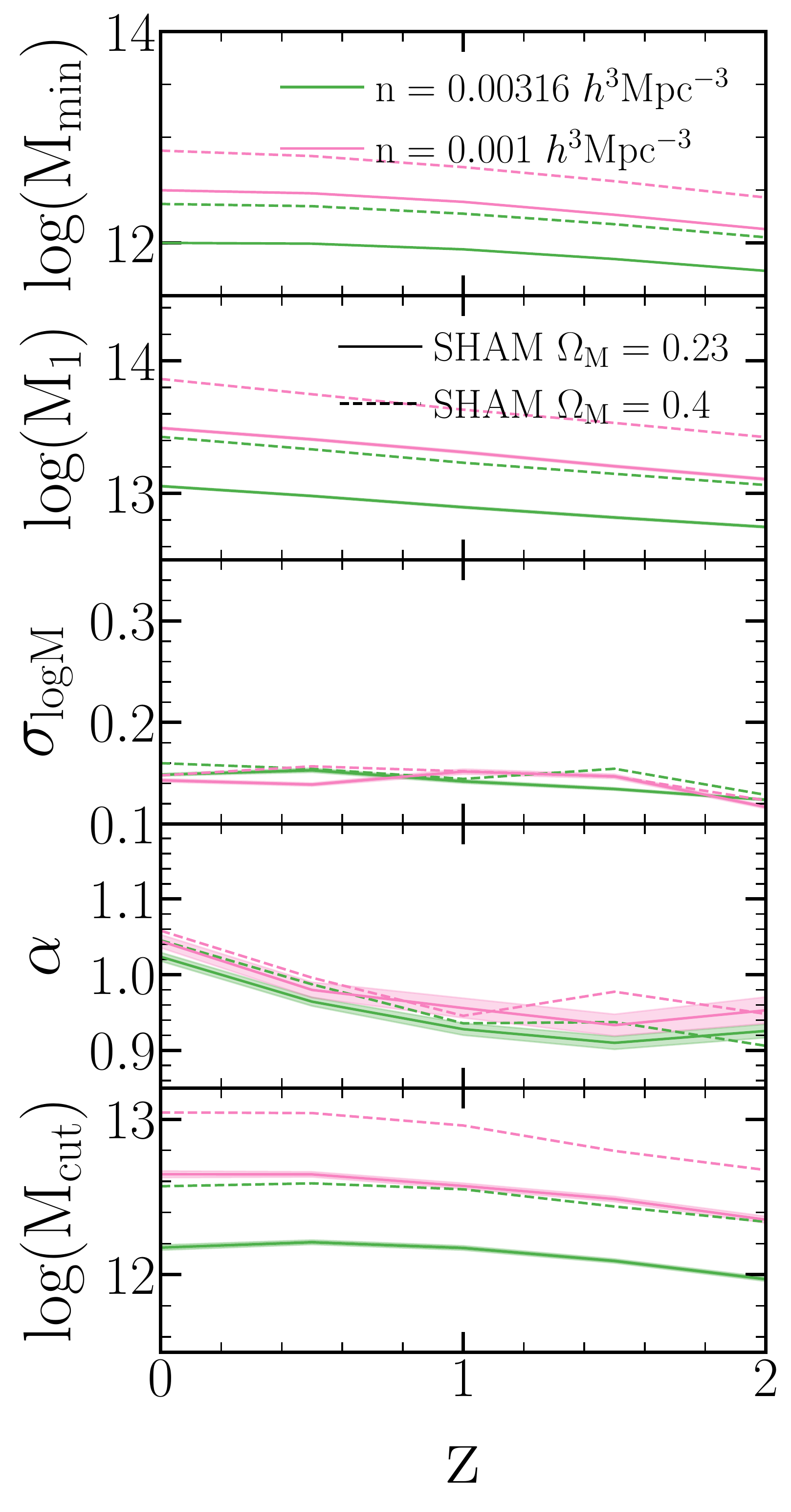}
\includegraphics[width=0.24\textwidth]{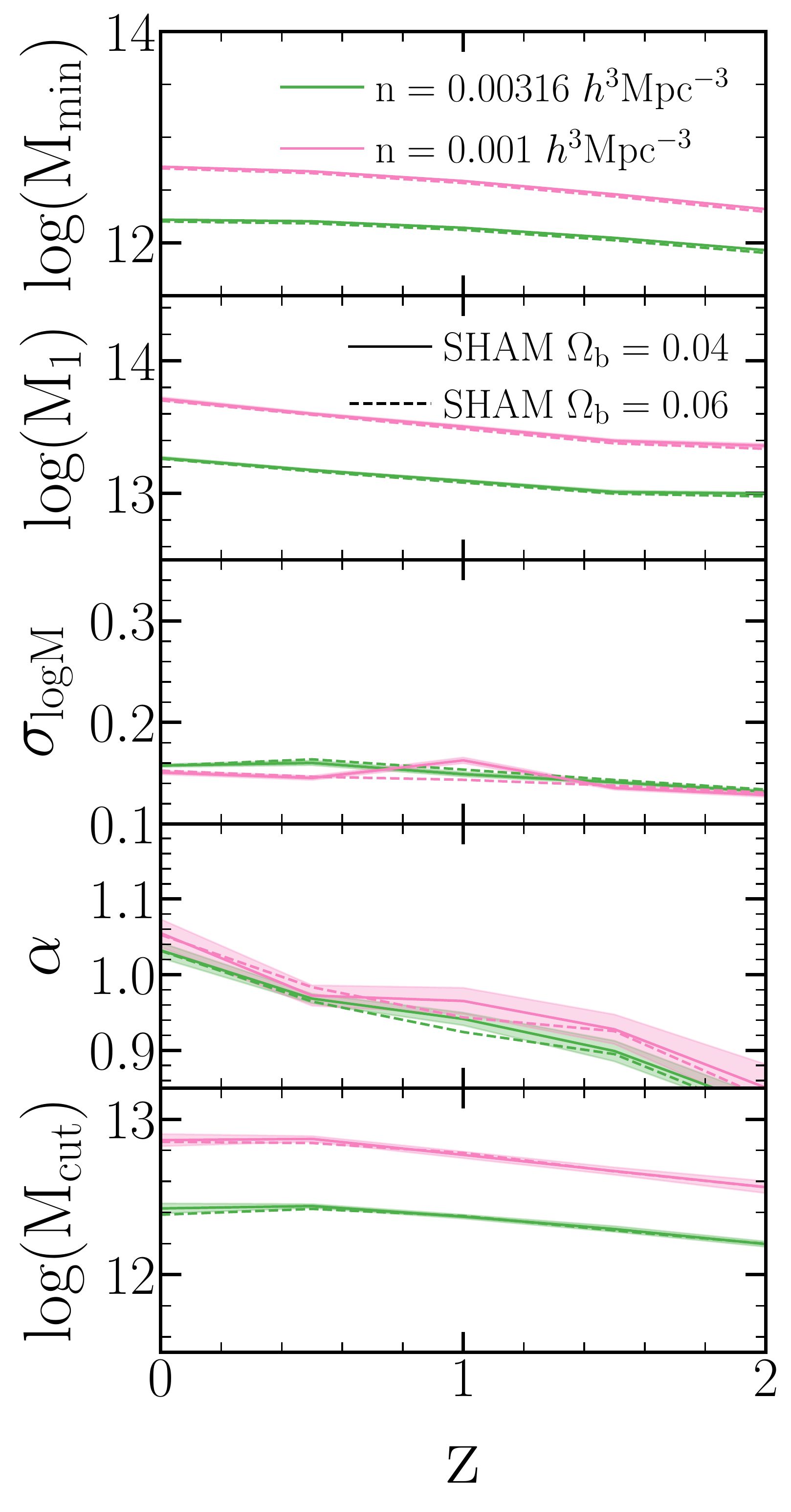}
\includegraphics[width=0.24\textwidth]{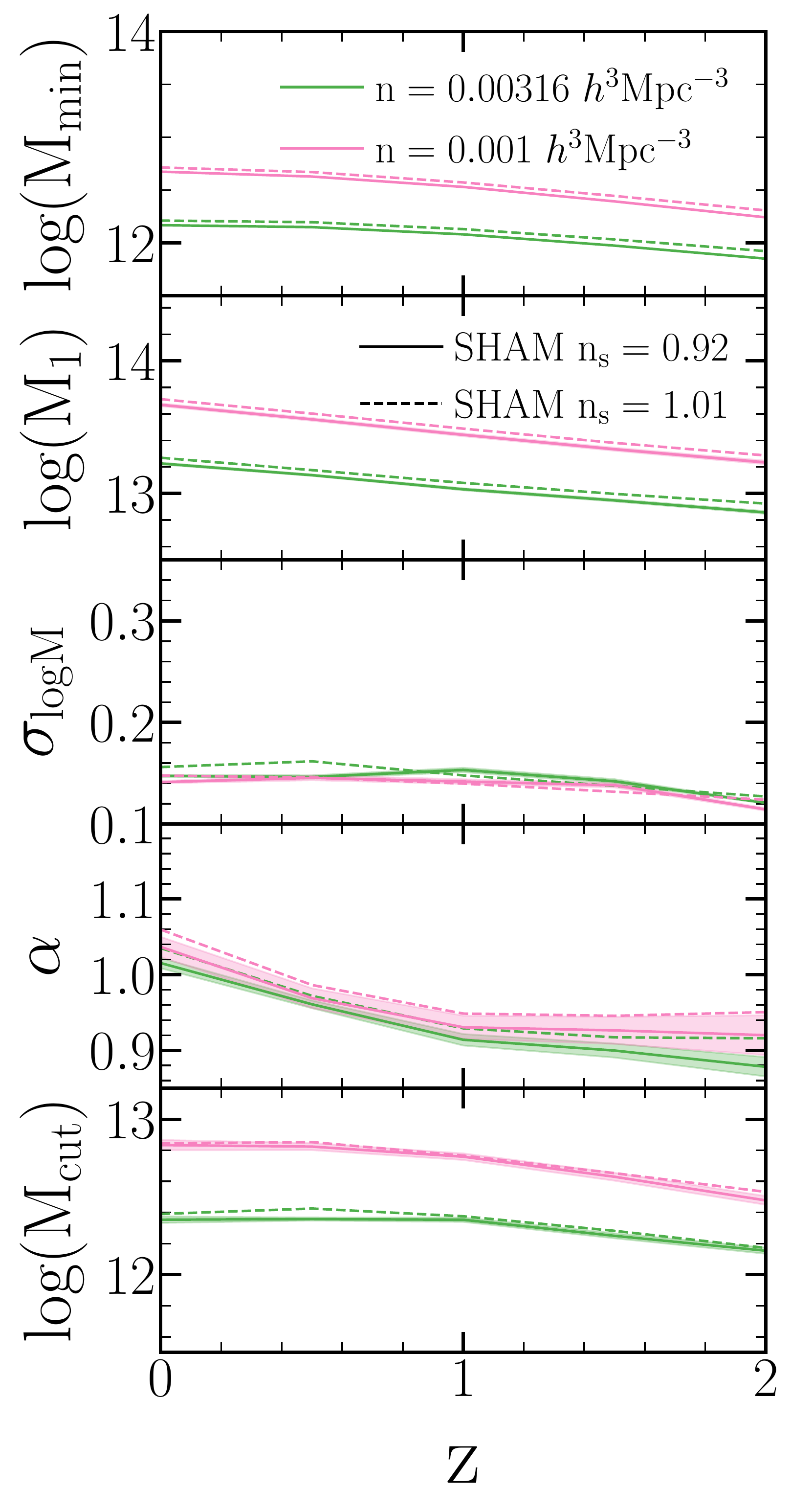}
\includegraphics[width=0.24\textwidth]{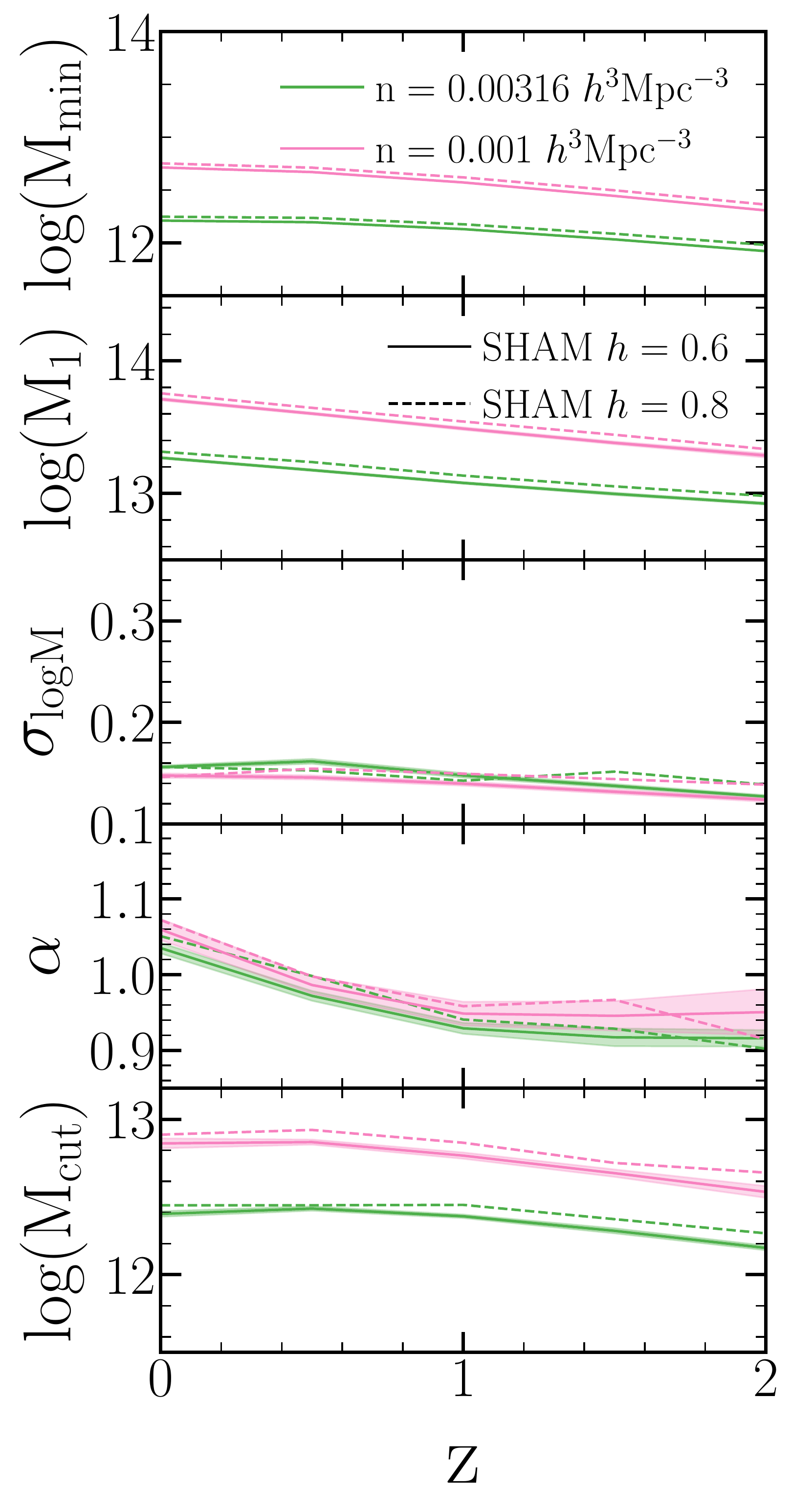}
\includegraphics[width=0.24\textwidth]{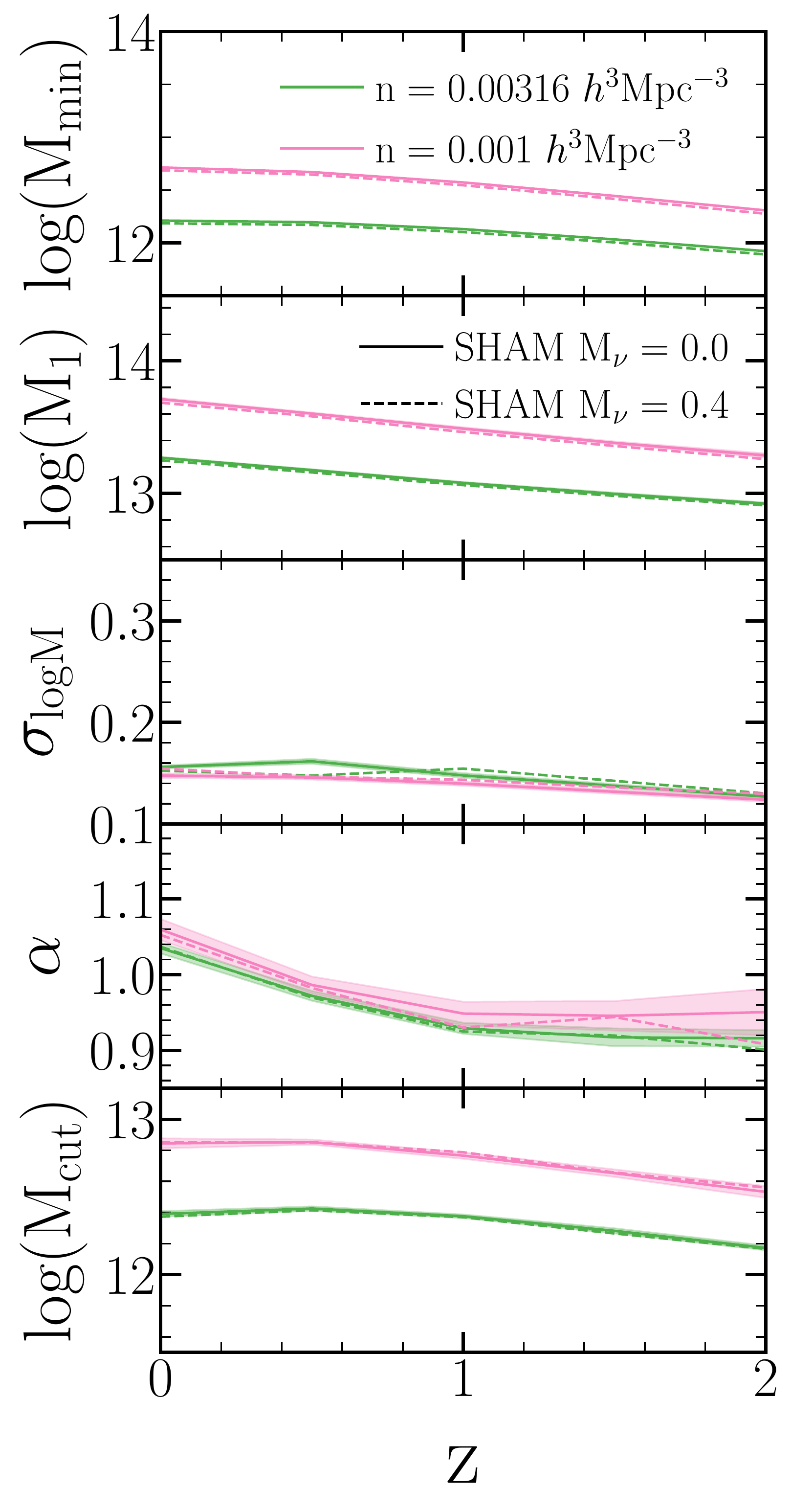}
\includegraphics[width=0.24\textwidth]{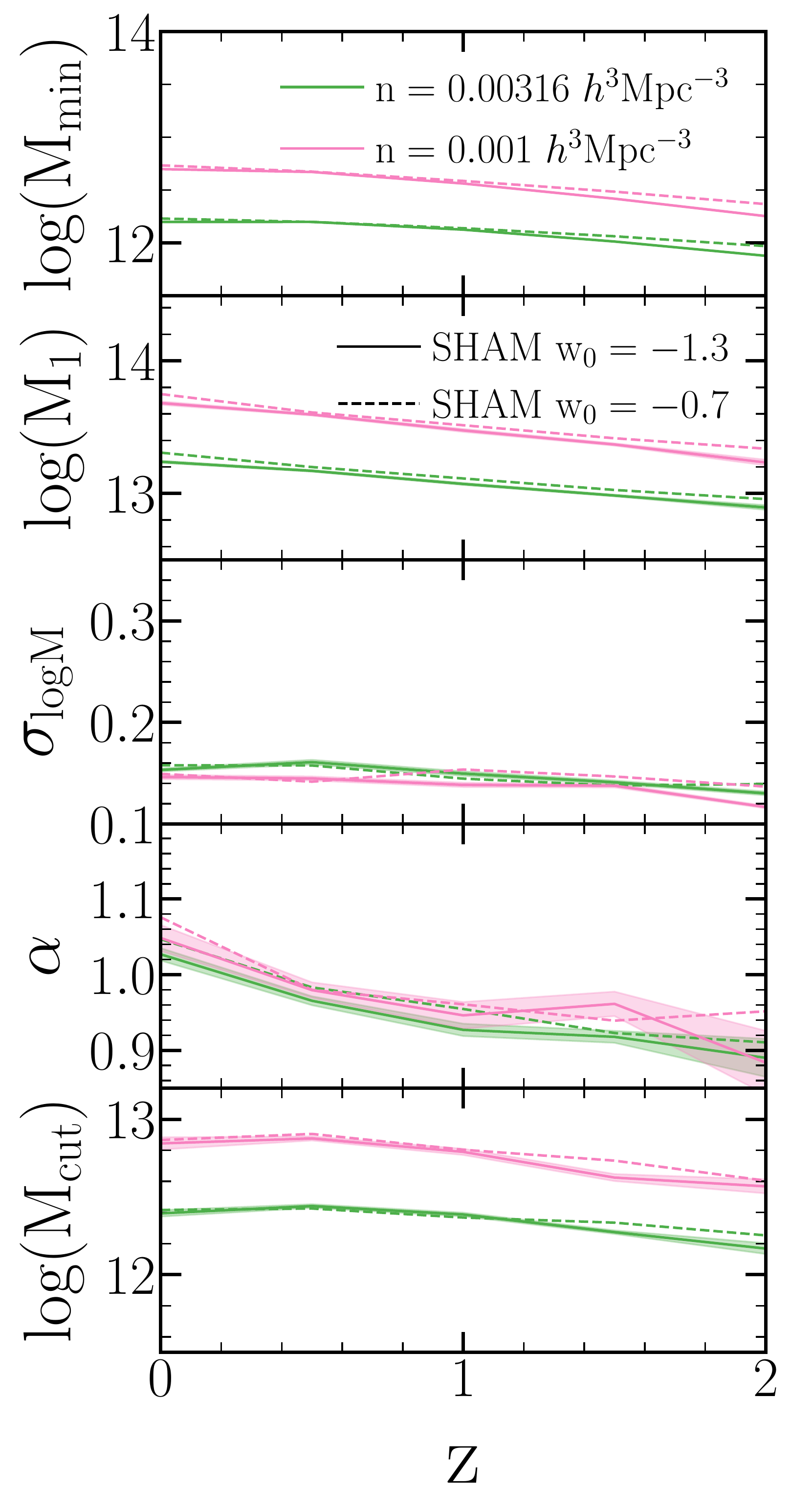}
\includegraphics[width=0.24\textwidth]{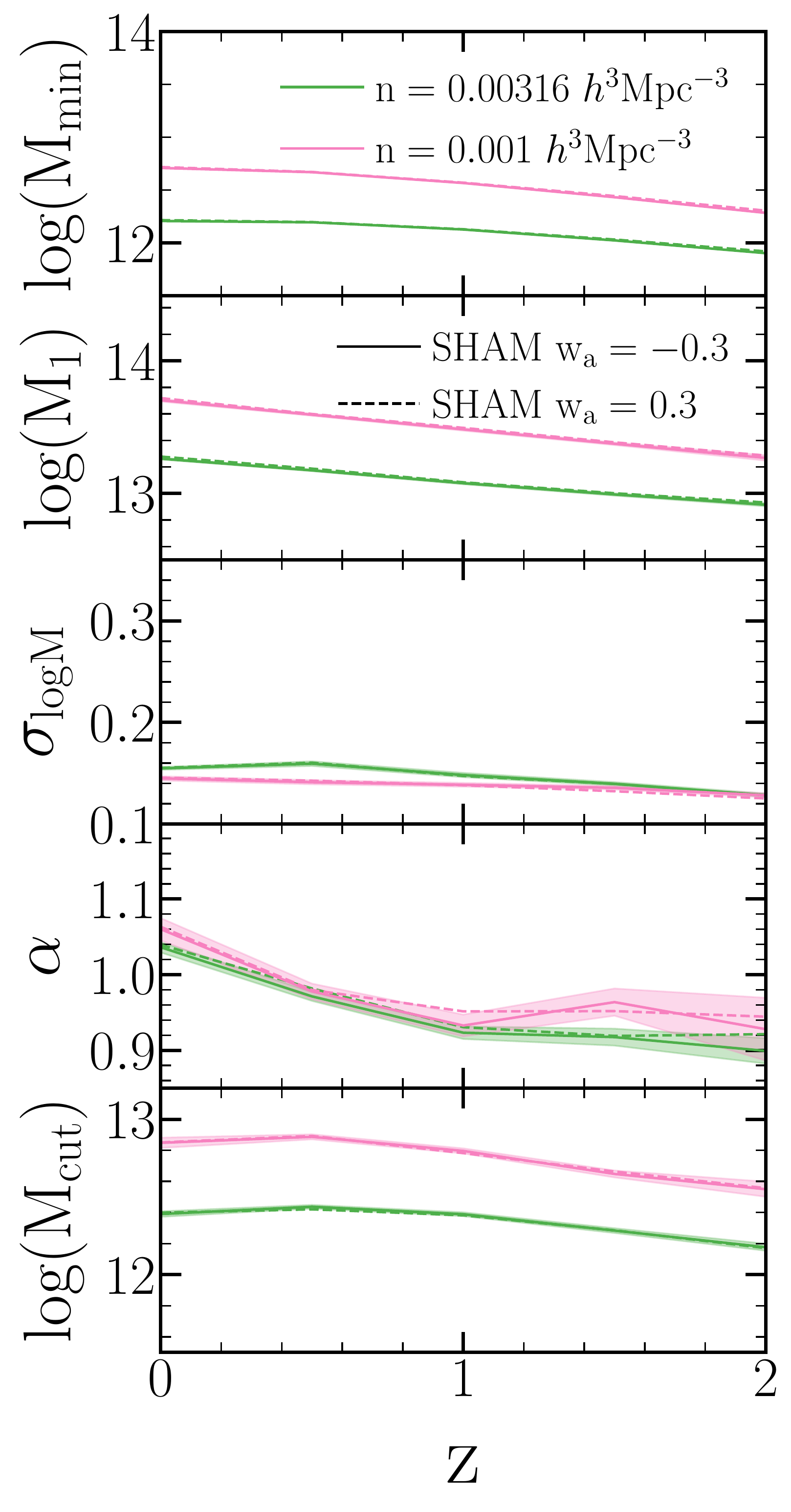}
\caption{Redshift evolution of the HOD parameters for the SHAM model without scatter, run on simulations with different values of the cosmological parameters. In each panel, the SHAM mocks were run on two dark matter-only simulations with identical cosmologies except for a single cosmological parameter. The cosmological parameters we change are (from left to right and top to bottom) $\sig$, $\OmM$, $\Omb$, $\ns$, $h$, $\Mnu$, $\wz$ and $\wa$. These changes represent significant variations in cosmology (about $\pm10\sigma$ around the Planck best-fits in most cases). See further discussion in the text.}  
\label{Fig:Cosmo}
\end{figure*}
 
\begin{table}
    \centering
    \caption{Cosmological parameters of the 13 pairs of simulations used in this work, spanning a $\sim 10\sigma$ parameter space around the best-fitting values of Planck. The top row specifies the cosmological parameters of the base cosmology. Each parameter is varied in turn as denoted in bold font in each subsequent line (based on Table~1 of \citealt{C21b}; see text for more details).}
    \begin{tabular}{cccccccc}
        \hline
        $\OmM$ & $\Omb$ & $\h$ & $\ns$ & $\sig$ & $\Mnu$ [eV] & $\wz$ & $\wa$\\
        \hline
        0.265 & 0.050 & 0.60 & 1.01 & 0.9 & 0 & -1 & 0 \\
        \hline
        0.265 & 0.050 & 0.60 & 1.01 & {\bf 0.73} & 0 & -1 & 0 \\
        {\bf 0.23} & 0.050 & 0.60 & 1.01 & 0.9 & 0 & -1 & 0 \\
        {\bf 0.4} & 0.050 & 0.60 & 1.01 & 0.9 & 0 & -1 & 0 \\
        0.265 & {\bf 0.040} & 0.60 & 1.01 & 0.9 & 0 & -1 & 0 \\
        0.265 & {\bf 0.060} & 0.60 & 1.01 & 0.9 & 0 & -1 & 0 \\
        0.265 & 0.050 & 0.60 & {\bf 0.92} & 0.9 & 0 & -1 & 0 \\
        0.265 & 0.050 & {\bf 0.80} & 1.01 & 0.9 & 0 & -1 & 0 \\
        0.265 & 0.050 & 0.60 & 1.01 & 0.9 & {\bf 0.4} & -1 & 0 \\
        0.265 & 0.050 & 0.60 & 1.01 & 0.9 & 0 & {\bf -1.3} & 0 \\
        0.265 & 0.050 & 0.60 & 1.01 & 0.9 & 0 & {\bf -0.7} & 0 \\
        0.265 & 0.050 & 0.60 & 1.01 & 0.9 & 0 & -1 & {\bf -0.3} \\
        0.265 & 0.050 & 0.60 & 1.01 & 0.9 & 0 & -1 & {\bf 0.3} \\
        \hline
    \end{tabular}
    \label{tab:cosmologies}
\end{table}

To look at the impact of cosmology on the evolution of the HOD, we populate galaxies using the SHAM technique in simulations of dark matter with varying cosmologies. Specifically, we employ a suite of 26 simulations utilised by \citet{C21b}, which is a subset of the {\tt BACCO} suite of simulations \citep{C20, Angulo:2021}. Here, we provide a brief summary of the most important features of these simulations; for more information, please refer to the papers listed above.

Each simulation was carried out with an updated version of {\tt L-Gadget3} \citep{Angulo:2012}, which is a lean version of {\tt GADGET} \citep{Springel:2005}, used to run the Millennium XXL simulation and the Bacco Simulations \citep{Angulo:2021}. For each cosmology, we ran two simulations with the same fixed initial density amplitudes but inverted phases, using the ``Fixed \& Paired'' technique, which allows us to suppress cosmic variance by at least 2 orders of magnitude on scales $k < 0.1 \ihMpc$ \citep{Angulo:2016}. All simulations have $1536^3$ dark matter particles in a $\sim (512\ \hMpc)^3$ volume. This is a significantly higher volume but lower resolution than the TNG300 and the TNG300-Dark. To mitigate resolution effects on our results, we limited ourselves to studying only the two lowest number density samples used in this paper.

The cosmological parameters of all simulations used are specified in Table~\ref{tab:cosmologies}. The cosmologies of the 13 pairs of simulations were originally chosen to study the performance of the scaling technique \citep{Angulo:2010, C20}. They are all variations of one of the main cosmologies of the {\tt BACCO} project, ``Nenya''. The other simulations have the same cosmology as Nenya except for one cosmological parameter which is varied, so that we can compare the results for two values of each cosmological parameter (e.g., $\sigma_8 = 0.73$ and $0.9$). The change in each cosmological parameter is considerable: about 10$\sigma$ variation around the Planck cosmology for the amplitude of density fluctuations $\sig$, the matter density $\OmM$, the baryon density $\Omb$ \& the spectral index $\ns$; more than $10\sigma$ for the Hubble parameter $h$ (so that it can cover SN predictions); about $5\sigma$ around the Planck best-fit for the neutrino masses $\Mnu$ and for the dark energy equation of state parameter $\wz$ and roughly $1\sigma$ for $\wa$ \citep{Planck:2018}.

We repeat the full SHAM analysis for each of our simulations, building a galaxy catalogue based on the values of $\vpeak$ of the subhaloes. We then compute the HOD at the different redshifts and number densities and express the evolution of the HOD as the evolution of its fitting parameters. Since we have two simulations for each cosmology, we use the mean for each parameter. 

The evolution of the HOD parameters is presented in Fig.~\ref{Fig:Cosmo}, where each set of panels shows the variation due to one cosmological parameter. There are noticeable differences between the values of the HOD parameters for some cosmologies (such as $\OmM$), but most of these differences are caused by the dependence of the halo mass function on cosmology. When focusing only on the evolution of the HOD parameters (and not on the value of the parameters at fixed redshift), we see that all the parameters evolve very similarly, namely that they follow the functional form presented in \S~\ref{subsec:HOD} (originally presented in \citetalias{C17}). The largest differences in the redshift evolution of the HOD parameters occur when varying $\sig$ and $\wz$ (although keeping the functional form proposed in this work). The changes are expected since these cosmological parameters are related to the way structures grow as a function of redshift.  Still, these changes in the evolution of the HOD parameters are relatively minor, especially considering the large variation in cosmology between the simulations. 

  These results indicate that any cosmological model within the $\rm  \Lambda CDM$ (extended) framework has a universal form of how the HOD parameters evolve. This conclusion, combined with the main results of this work showing that this same redshift evolution of the HOD parameter is valid for a variety of galaxy formation models, provides a solid ansatz when creating HOD mock catalogues in simulated lightcones. Namely, that the functional form assumed should be equally valid independent of the galaxy formation model or the cosmological parameters of a $\rm  \Lambda CDM$ model (and most probably, any model where structure grows hierarchically).


\bsp	
\label{lastpage}
\end{document}